\documentclass[aps,prx,twocolumn,groupedaddress]{revtex4-2}
\usepackage{graphicx}
\usepackage{microtype}
\usepackage[dvipsnames]{xcolor}
\usepackage{tabularx}
\usepackage[colorlinks=true,linkcolor=blue,allcolors=blue]{hyperref}%

\begin{document}


\title{Soft-x-ray spectroscopy  of two-dimensional transition metal dichalcogenides: valley selectivity?}

\author{Andrey Geondzhian}
\affiliation{Max Planck Institute for Structure and Dynamics of Matter, 22761 Hamburg, Germany}
\affiliation{Max Planck POSTECH/KOREA Research Initiative, 37673 Pohang, South Korea}
\author{Angel Rubio}
\affiliation{Max Planck Institute for the Structure and Dynamics of Matter, 22761 Hamburg, Germany}
\affiliation{Center for Computational Quantum Physics (CCQ), The Flatiron Institute, 162 Fifth Avenue, New York, NY 10010, USA}
\author{Massimo Altarelli}
\email{massimo.altarelli@mpsd.mpg.de}
\affiliation{Max Planck Institute for the Structure and Dynamics of Matter, 22761 Hamburg, Germany}

\date{\today}							
\begin{abstract}
Optical properties of semiconducting monolayer transition metal dichalcogenides have received a lot of attention in recent years, following the discovery of the valley selective optical population of either $K_{+}$ or $ K_{-}$ valleys at the direct band gap, depending on the polarisation of the incoming light. We use group theoretical selection rules, as well as \textit{ab initio} DFT calculations, to investigate whether this valley selectivity effect is also present in x-ray optical transitions from the flat core level of the transition metal atom to the valence and conduction band $K$ valleys. Valley selectivity is predicted for $s$, $p_{1/2}$ and $p_{3/2}$ edges in transitions to and from the valence band edges with circularly polarised radiation. Possible novel applications to the diagnostics of valleytronic properties and intervalley dynamics are investigated and the feasibility of ultrafast pump-probe and Kerr rotations experiments with suitable soft-x-ray free-electron laser sources is discussed.
\end{abstract}

\maketitle
\section{Introduction\label{Intro}}
Two-dimensional transition metal dichalcogenides (2D-TMD) provide a very interesting platform for the investigation of two-dimensional solids, with some specific features that differentiate them from either graphene or $hBN$~\citep{HeinzRMP}. In particular, the semiconducting members of this family ($MX_2$ compounds, with $M=Mo, W$ and $X=S, Se, Te$) display interesting optical properties in the fundamental gap region; the monolayer gap is direct, and located at two distinct opposite corners of the Brillouin zone, denoted in the following by $K_{+}, K_{-}$, referred to as "valleys", where pronounced exciton effects are observed and where selective optical population of either valley, depending on the sign of the circularly polarised incoming light~\cite{Cao2012, Xiao2012}, is observed. This effect has led to investigate the potential use of these \textit{"valleytronics"} effects for advanced technological applications. The origin of the effect in 2D-TMD and other compounds with the same symmetry is ascribed to the lack of inversion symmetry and to the mirroring of the $K_{+}, K_{-}$ band extrema under time-reversal symmetry \cite{Vitale}. The role of the spin-orbit interaction, frequently emphasized in the literature, on the other hand, is not essential (see Appendix A for a numerical example).

One of the hotly discussed topics is the understanding and possible control of the decay time of the difference in occupation of the two valleys, related to the dynamics of electrons, holes and excitons in interaction with the other elementary excitations of the two-dimensional crystal. In this context, it may be of interest to analyse alternative excitation paths, that for example do not simultaneously generate valence band holes and conduction  electrons in the selected band valley. 

Here we consider the possibility to excite electrons from core levels by absorption of x-ray photons, or to fill core holes by valence electrons in x-ray emission processes, investigating possible valley selectivity effects of the photon polarisation. Given that the conduction and valence band valleys have more than 75\%  metal $d$-states character~\cite{Cappelluti}, it is to be expected that the most intense dipole allowed transitions occur from the $p$-like core levels of the transition metal atoms, for example at the $L_{2,3}$, $M_{2,3}$, or $N_{2,3}$ edges, corresponding to $2p$, $3p$, $4p$ core levels (although the low, parity-breaking lattice symmetry allows other possibilities, as we shall soon see). 

Very recently, ultrafast far-UV and soft-x-ray spectroscopy investigations of transition metal dichalcogenides films, after excitation with a pump laser (with duration $<5~\rm{fs}$), have been performed using high harmonic generation sources~\cite{Attar, Chang}, and a free-electron laser~\cite{Britz}, with a pump duration $\simeq 50~\rm{fs}$; the samples, however, were multilayer films with several tens of $\rm{nm}$ thickness, and the  photons were not circularly polarised. These pioneering experiments were therefore addressing interesting issues of carrier dynamics, but, because of the sample and photon source properties, were not accessing valleytronic properties, which are the focus of the present work.

Practical reasons suggest to concentrate attention on rather shallow core levels, with ionisation energies restricted to the far UV or soft-x-ray regions. Spectroscopy with photon energies of several $keV$ can only be performed with limited energy resolution, unsuitable to differentiate band features such as primary and secondary valleys; in addition the photon wavevector at such energies cannot be considered as negligible with respect to the size of the Brillouin zone, and therefore optical transitions cannot be described in the dipole approximation, with initial and final single-electron states with the same quasi momentum in the zone. 

This paper is organised as follows. In Section 2 standard group-theoretical techniques are applied to 2D-TMD, recovering the known symmetry characters of valence and conduction band edges \cite{SongDery} , but also identifying the irreducible representations of the $s-$, $p-$ and $d-$like core level edges and deriving the selection rules for optical transitions, for both core and valence excitations. The results, pointing to the valley selectivity of circularly polarised x-ray transitions to the valence bands from the $s$ and $p$ core level edges, but not from the $d$ edges, are briefly discussed. Transitions into valence band states, of course, require also that the Fermi level is below the valence band top, which can occur by electrostatic gating, by optical pumping or by p-type doping. In Section 3, \textit{ab initio} DFT calculations of the x-ray absorption spectra are presented, that confirm and complement the group theoretical results. In Section 4, taking inspiration from existing experimental results for multilayer TMD, we discuss possible contributions of laser pump, soft-x-ray probe spectroscopy to the study of the electronic properties and dynamics of valley-pumped systems. The requirements on the soft-x-ray source (high intensity and monochromaticity, short pulses and variable polarisation) point in the direction of seeded free-electron lasers~\cite{Fermi}. 

\section{Group-theoretical selection rules for core electron transitions }\label{sec:GT}

Van der Waals multilayer stackings of transition metal dichalcogenides, $MX_{2}$, have been investigated for many years. The progress in the manipulation and exfoliation of two-dimensional structures lead more recently to the fabrication of monolayers of the 2H structure that, unlike the multilayer versions, have a direct band gap~\cite{HeinzRMP}. In Figure 1(a) and (b) a schematic side and top view of the monolayer structure are shown. It consists of three planes of a triangular arrangement of atoms, with the metal M atoms at the central plane, sandwiched between the two chalcogen X planes, with a stacking that locates metal and chalcogen atoms, when projected on a single plane, at alternating vertices of regular hexagons. If we denote by $\it{a}$ the distance of two consecutive vertices of a hexagon (Figure 1(b)) $\it{a}$ is \textit{e.g.} approximately $0.182\,nm$ for $MoS_2$ and $0.189\,nm$ for $WSe_2$, as can be derived from the crystallographic data reproduced by Korm{\'{a}}nyos \textit{et al.}~\cite{Kormanyos2015}. In terms of this $\it{a}$ length, the cartesian coordinates of the lattice unit vectors $\mathbf{a_1, a_2}$ in terms of the $x,y$ axes of Figure 1(b) are
\begin{equation}
\mathbf{a_1} = \frac{a}{2} (3,\sqrt3),\; \; \; \mathbf{a_2} = \frac{a}{2}(3,-\sqrt3).
\end{equation}
%

\begin{figure*}[t]
\includegraphics[width=15cm]{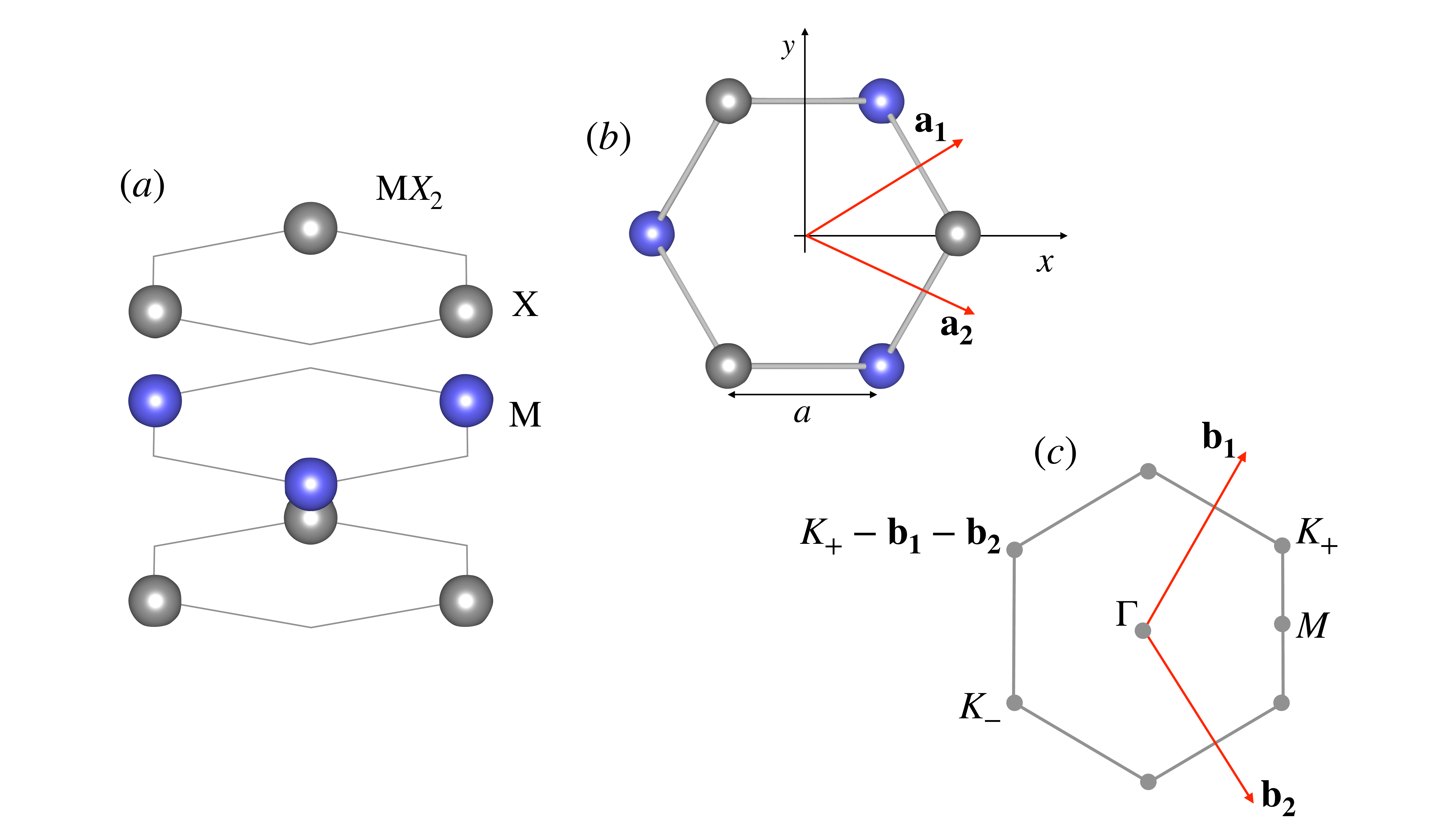}
\caption{Structure of  MTMDs. (a) shows a perspective side view, (b) a top view, including the two unit vectors $\mathbf{a_1}$, $\mathbf{a_2}$ and an orthogonal cartesian frame to describe all vectors in real and reciprocal space. (c) Brillouin zone for the two-dimensional lattice structure, with the unit reciprocal lattice vectors $\mathbf{b_1}$, $\mathbf{b_2}$ and with the two wavevectors of the band structure "valleys", $\mathbf{K_+}$, $\mathbf{K_-}$. A vector equivalent to $\mathbf{K_+}$ is shown for illustration purposes.}
\end{figure*}

In Figure 1(c) the corresponding Brillouin zone (BZ) is shown, with the unit vectors of the reciprocal lattice $\mathbf{b_1}, \mathbf{b_2}$ and with $\mathbf{K_+},  \mathbf{K_-}$ as the two opposite, non-equivalent points where the direct gap is located. It is important to note that, in the same cartesian frame
\begin{equation}
\mathbf{b_1}= \frac{2\pi}{3a}(1,\sqrt{3}), \; \;\; \mathbf{b_2}= \frac{2\pi}{3a}(1,-\sqrt{3}) ,
\end{equation}
and the $K_+$ and $K_-$ points are not connected by a reciprocal lattice vector
\begin{equation}
 \mathbf{K_+} = \frac{2\pi}{3a}(1, \sqrt{3}/3) = -  \mathbf{K_-}  \ .
 \end{equation}
 %
Also, the star of  $\mathbf{K_+}$ only contains (besides itself) $\mathbf{K_-}$ because all other vertices of the hexagonal BZ are connected to either $\mathbf{K_+}$ or $\mathbf{K_-}$ by a reciprocal lattice vector (an example is shown in Figure 1(c)). It is important to remember that, since in the case of the 2D-TMD inversion is absent, Kramers' theorem \cite{Kittel} does not apply, and non-degenerate spin bands occur throughout the zone. Furthermore, the point group of the structure shown in Figure 1 is the dihedral group $D_{3h}$.
To determine the symmetry properties and the selection rules for optical transitions of the Bloch functions at the $\mathbf{K_+}$ or at the $\mathbf{K_-}$ point, we have to consider the "small group" or "wavevector group" at this point, the subgroup of $D_{3h}$ containing the operations that leave this point invariant \cite{Tinkham, BraCra}. It easy to see that the $C_2$ rotations and the corresponding $\sigma_v$ mirror reflections of  $D_{3h}$ interchange the two (non-equivalent) vectors, and are therefore not  part of the small group. The remaining operations form the point group $C_{3h}$. This is a rather simple abelian group, with six operations and six classes, each comprising one element
\begin{equation}
E \;\;\;\;\; C_{3}^+ \;\;\;\;\; C_{3}^- \;\;\;\;\; \sigma_{h} \;\;\;\;\; S_{3}^+ \;\;\;\;\; S_{3}^-,
\end{equation} 
where $E$ is the identity, the $2\pi/3$ rotations $C_{3}$ and the rotation-reflections $S_{3}$ (about an axis perpendicular to the layer and intersecting the centre of a hexagon) are labelled $+$ or $-$ depending on whether they are counterclockwise or clockwise. To obtain all the required information about the representations of the point group $C_{3h}$ and of other groups we shall need later, such as the full rotation-reflection group in 3-dimensional space, $O(3)$, many textbooks can be consulted. In the following the notations and conventions of the excellent compilation by Koster, Dimmock, Wheeler and Statz~\cite{KDWS} (KDWS) shall be adopted.

In order to classify the symmetry of electronic states at the relevant wavevectors $\mathbf{K_+}$ or $\mathbf{K_-}$ we must first determine the action of the six $C_{3h}$ operations on the corresponding Bloch functions. We shall suitably adapt the treatment, as given in ref.~\cite{SongDery, Kormanyos2015, Ochoa}, for the lowest conduction and upmost valence bands to include the case of bands derived from core levels. 

\subsection{Symmetry character of conduction and valence levels at \texorpdfstring{$\mathbf{K_+}, \mathbf{K_-}$}{Lg}}

To establish the basic methods, notations and conventions, we sketch here the derivation of the symmetry properties and optical selection rules for the valence to conduction band transitions (for more details, see Appendix B).
 
Following the results of \textit{ab initio} and semi-empirical band structure calculations, the dominant orbital character of lowest conduction band valleys at $\mathbf{K_+}$ and  $\mathbf{K_-}$ is from the $m=0\; d-$orbital of the metal atom (with a weight e.g. for $MoS_2$ estimated~\cite{Cappelluti} at 82\%). The remaining fraction is ascribed to $p$ orbitals of the chalcogen, but for symmetry considerations this adds no relevant information. The Bloch function at $\mathbf{K_\pm}$ can be written in the general form~\cite{Kormanyos2}
\begin{equation}
\Psi_{c, \mathbf{K_\pm}} = \sum_{j_{1},j_{2}} e^{i\mathbf{K_\pm}\cdot \mathbf{R}_{j_{1},j_{2}}} \Phi_{2,0}(\mathbf{r-R}_{j_{1},j_{2}})\chi(\sigma_{z}),
\end{equation}
where (see Figure 2) the metal atomic sites are identified by the integers $j_{1},j_{2}$ , according to
\begin{equation}
\mathbf{R}_{j_{1},j_{2}} = j_{1} \mathbf{a_1} + j_{2} \mathbf{a_2} + \mathbf{\delta_3},
\end{equation}
%
%
\begin{figure*}
\includegraphics[width=15cm]{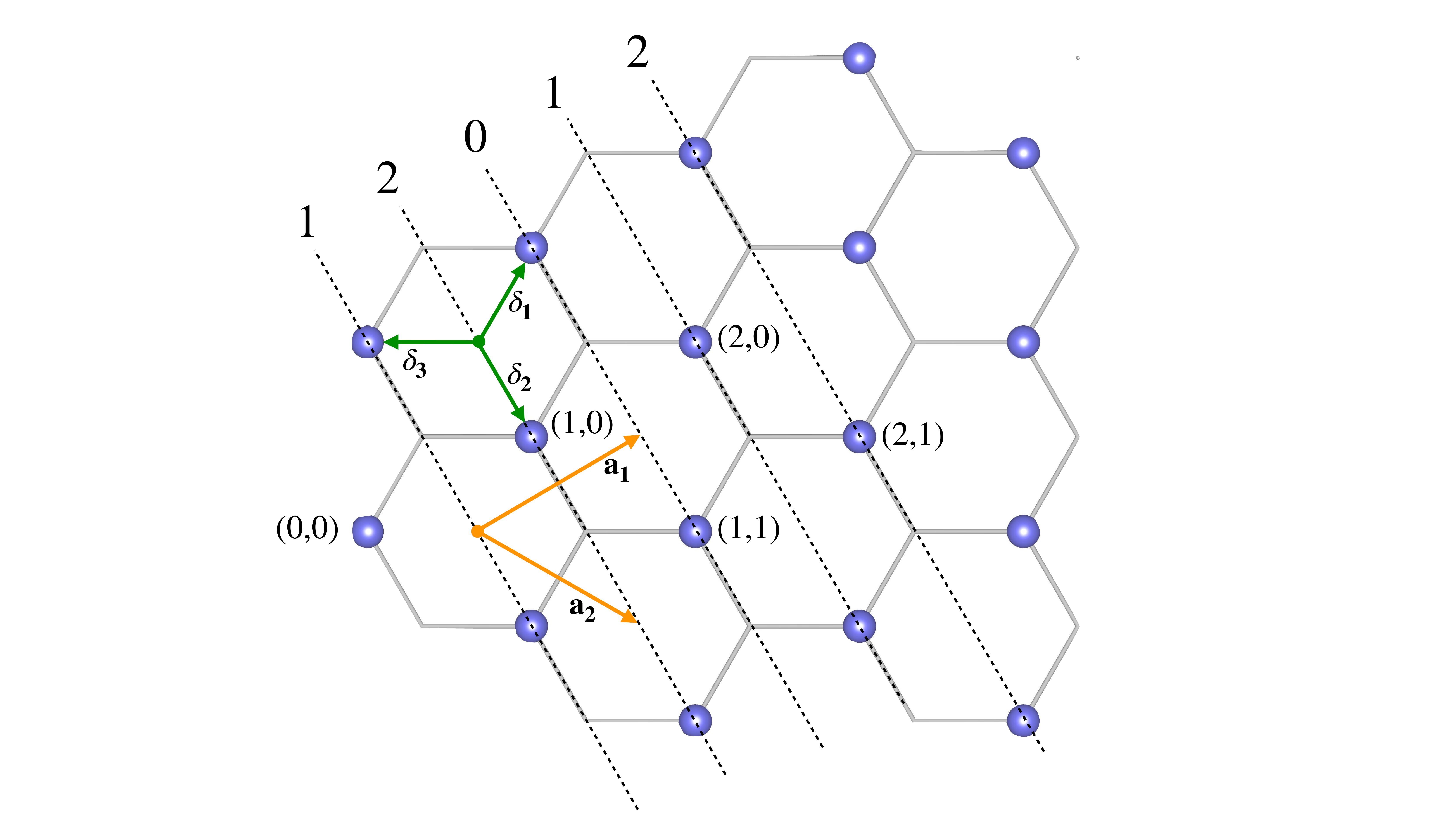}
\caption{The middle plane of the monolayer structure, showing the position of the metal atoms and the unit vectors $\mathbf{a_{1}, a_2}$. Some atoms are labelled with their  $j_{1}, j_2$ coordinates as an example. The three vectors $\mathbf{\delta_i}, i=1,2,3$ are also shown. The dotted diagonal lines are the wavefronts of the $\mathbf{K_+}$ Bloch waves, where $(2j_{1} + j_{2} - 1)$ is constant (see text). The integers at the top end of the dotted lines are $(2j_{1} + j_{2}\; mod \: 3).$}
\end{figure*}
%
%
%
having introduced the vectors
\begin{equation}
\mathbf{\delta_1}= \frac{a}{2} (1,\sqrt{3}),\;\;\; \mathbf{\delta_2}= \frac{a}{2} (1,-\sqrt{3}),\;\;\; \mathbf{\delta_3}= a (-1,0),
\end{equation}
and having shown in Figure 2, next to a few atoms, the $(j_{1}, j_{2})$ integer values as examples.

In each term of  Eq.~(5), the first factor is the Bloch phase factor, $\Psi_{2,0}$ is a Wannier function with $l=2, m=0$ character and $\chi(\sigma_{z})$ denotes the spinor wavefunction, distinguishing the spin up or down conduction valley. In adopting this form of the Bloch function to discuss the symmetry properties, we implicitly acknowledge that, in the fundamental gap region, the crystal field effects, that lift the fivefold degeneracy of the $d$ orbitals of the metal atoms by several $eV$'s, are more important than the spin orbit interaction, that mixes different $m$ and different $\sigma_z$ values.  To proceed, one must consider the effect of the $C_{3h}$ operations on each factor. In doing so, we follow the standard group-theoretical procedure~\cite{Tinkham, BraCra, KDWS} to describe the effect of a space symmetry operation $\hat{C}$ on a generic function $f$ of the space coordinates $\mathbf{r} \equiv (x,y,z)$ 
\begin{equation}
\hat{C} f(\mathbf{r}) = f(\hat{C}^{-1} \bf{r} ).
\end{equation}
As detailed in Appendix B, the Bloch phase factor at $K_+$ transforms according to the $\Gamma_3$ irreducible representation, and at $K_-$ according to $\Gamma_2$ ; the $d-$like Wannier function, on the other hand, acquires a factor $e^{-im\phi}$ for all rotations about the $z$ axis; so for $m=0$ it is invariant for all rotations and also for the perpendicular plane reflection, and therefore belongs to $\Gamma_1$ ~\cite{KDWS-t5759}. As to the spin wavefunctions $\chi$, they are by definition a basis for the two-dimensional $D_{1/2}^+$ representation of the full  rotation-reflection group O(3), where the $+$ superscript acknowledges that the spin pseudovector is parity even. It is important to recall that when dealing with half-integer spin systems we must consider the "double" groups, resulting from introducing an additional operation $\overline{E}$, that reverses the sign of the half-integer spin functions, but does not affect the space coordinates (equivalently, the additional elements of the group are often described by adding $2\pi$ to the angle of every rotation of the group)~\cite{Tinkham, BraCra, KDWS}; and consider only the so-called additional representations, odd under $\overline{E}$. When restricted to the operations of $C_{3h}$,  $D_{1/2}^+$ is reducible (the fact that $C_{3h}$ has six group elements (twelve for the double group) and six group classes (twelve for the double group) implies that all irreducible representations are one-dimensional, corresponding to the absence of spin degeneracy). We can reduce $D_{1/2}^+$ (as $C_{3h}$ representation) as follows~\cite{Table62}  
\begin{equation}
D_{1/2}^{+} = \Gamma_{7} + \Gamma_{8}.
\end{equation}
Therefore, from~\cite{KDWS-ch23} at $K_+$:
\begin{equation}
\Gamma_{3} \otimes D_{1/2}^{+}  = \Gamma_{3} \otimes \Gamma_{7} +\Gamma_{3} \otimes \Gamma_{8} = \Gamma_{10} + \Gamma_{12}
\end{equation}
Similarly, at $K_-$:
\begin{equation}
\Gamma_{2} \otimes D_{1/2}^{+} = \Gamma_{2} \otimes \Gamma_{7} +\Gamma_{2} \otimes \Gamma_{8} = \Gamma_{11} + \Gamma_{9},
\end{equation}
with the two representations corresponding to the two spin directions.

Note that $\Gamma_{9, 11}$ are respective complex conjugates of the $\Gamma_{10, 12}$. This is a reformulation of the time reversal symmetry condition  $E_{n}(\mathbf{k}, \uparrow) = E_{n}(-\mathbf{k}, \downarrow)$ implied by the discussion immediately following Eq.~(3). 

Consider now the top of the valence bands. The corresponding dominant orbital is the metal $d-$orbital with $m=\pm 2$ ~\cite{Kormanyos2}, schematically behaving as $(x\pm iy)^2$. We can write the Bloch functions as
\begin{equation}
\Psi_{v, \mathbf{K_\pm}} = \sum_{j_{1},j_{2}} e^{i\mathbf{K_\pm}\cdot \mathbf{R}_{j_{1},j_{2}}} \Phi_{2,\pm2}(\mathbf{r-R}_{j_{1},j_{2}})\chi(\sigma_{z}).
\end{equation}
In either case, namely  $K_+$, $m=+2$ or $K_-$, $m=-2$ it is readily seen that the product of the Bloch factor and the $d-$like Wannier function is invariant under all $C_{3h}$ operations~\cite{Cao2012}, belonging to the $\Gamma_1$ identical representation. The overall Bloch function with the spin part is a basis for the $D_{1/2}^+$, and we already know that  $D_{1/2}^+ = \Gamma_{7}  +\Gamma_8$. 

On the basis of this symmetry assignment, it is possible to establish the selection rules for optical transitions. The group theoretical necessary but not sufficient prescription for a non-vanishing transition matrix element from an initial state $\psi_{i}$ to a final state $\psi_{f}$ under the transition operator ($\hat{T}^{\alpha}$), 
\begin{equation} 
\hat{T}_{i,f}^{\alpha} = \langle \psi_{i}| \hat{T}^{\alpha}|\psi_{f}\rangle = \int d^{3}r \psi_{i}^{*} (\mathbf{r}) \hat{T}^{\alpha}\psi_{f}(\mathbf{r}),
\end{equation}
 can be expressed as~\cite{Tinkham-s4}
\begin{equation}
\Gamma_{\alpha} \otimes \Gamma_{f} = \Gamma_{i} .
\end{equation}
The fact that the "small" group of the wavevectors  $\mathbf{K_+}, \mathbf{K_-}$ has only one-dimensional irreducible representations makes our life much easier (do not need to specify and discuss individual rows of irreducible representations). For the valence and conduction edges we determined the representations; for the transition operator $\mathbf{A\cdot p}$, corresponding to dipole optical transitions, we have the vector potential
\begin{equation}
\mathbf{A\cdot p} \sim \sum_{\mathbf{k}, \lambda} [\mathbf{\epsilon}_{\lambda} b_{\mathbf{k}, \lambda} e^{i \mathbf{k \cdot r}} + \mathbf{\epsilon^*}_{\lambda} b_{\mathbf{k}, \lambda}^{+} e^{-i \mathbf{k \cdot r}}] \cdot \mathbf{p},
\end{equation}
where $\mathbf{\epsilon}$ denotes the polarisation vector of mode $\lambda= \pm1, 0$ for positive circular, negative circular and linear $z$ polarisation respectively,  and $b^{+}, b$ are photon creation and annihilation operators. Including also the photon states, we can precisely define the transition operator in Eq.~(15) for a photon absorption process  as
\begin{widetext}
\begin{equation} 
\langle \psi_{i}; n_{\lambda}| \hat{T}^{\alpha}|\psi_{f}; n_{\lambda}-1\rangle = \langle \psi_{i}; n_{\lambda}| [\mathbf{\epsilon^*}_{\lambda} b_{\mathbf{k}, \lambda}^{+} e^{-i \mathbf{k \cdot r}}] \cdot \mathbf{p}|\psi_{f}; n_{\lambda}-1\rangle,
\end{equation}
\end{widetext}
with $n_{\lambda}, n_{\lambda}-1$ denoting the number of photons of the $\lambda$ polarization in the initial and final state respectively.

The transition operator (in the $\mathbf{k} \longrightarrow 0 $ limit) contains therefore  $\mathbf{\epsilon}_{\lambda}\cdot \mathbf{p}$ (for photon emission) and $\mathbf{\epsilon^{*}}_{\lambda}\cdot \mathbf{p}$ (for photon absorption), therefore transforms like the components of the $\mathbf{p}$ vector, i.e, like the $\mathit{l} = 1$ spherical harmonics, or the $D_{1}^{-}$ representation of $O(3)$. The reduction of $D_{1}^{-}$ in $C_{3h}$ is~\cite{KDWS-t58}
\begin{equation}
D_{1}^{-} = \Gamma_{2} +  \Gamma_{3} +  \Gamma_{4}
\end{equation}
and from a look at the character table of  $C_{3h}$, keeping in mind the convention embodied in Eq.~(8) and applying it to $\mathit{l} = 1$ spherical harmonics~\cite{Tinkham-p64}, it is easy to establish that
\begin{eqnarray}
\epsilon_{+1}, \epsilon^{*}_{-1}& \sim  (p_{x}+ip_{y})  \longrightarrow  \Gamma_{3},  \nonumber \\ 
\epsilon_{-1}, \epsilon^{*}_{+1}&\sim (p_{x}-ip_{y})  \longrightarrow  \Gamma_{2}, \nonumber  \\
\epsilon_{0}   &\sim p_{z} \longrightarrow \Gamma_{4},
\end{eqnarray}
thus assigning the proper symmetry to positive and negative circular polarisation for propagation perpendicular to the plane, and to linear polarisation perpendicular to the plane. We are especially concerned with the first two, that determine the chirality of optical properties of MTMD. Taking Eq.~(16) into account, we identify $\Gamma_2$, ($\hat{T}^{\alpha} \sim \epsilon^{*}_{+1}$) as the $'+'$ circular polarisation, and $\Gamma_3$ ($\hat{T}^{\alpha} \sim \epsilon^{*}_{-1}$) as the $'-'$ one, following the convention implicitly assumed in ref.~\cite{Cao2012}.

A straightforward perusal of the $C_{3h}$ table of multiplication~\cite{KDWS-t58} and character table~\cite{KDWS-t57},  yields the following allowed transitions from the valence band state $\Gamma_{7}, \Gamma_{8}$, for circularly polarized positive (+) or negative (-) photons.

At $\mathbf{K_+}$:
\begin{displaymath}
\begin{array}{lcr}

Initial &     Final &    Polarization \\

\Gamma_{7} & \Gamma_{10} &  + \\

\Gamma_{8} & \Gamma_{12} &  +

\end{array}
\end{displaymath}

At $\mathbf{K_-}$:

\begin{displaymath}
\begin{array}{lcr}

Initial &     Final &    Polarization \\
\Gamma_{7} & \Gamma_{11} &  - \\

\Gamma_{8} & \Gamma_{9} &  -

\end{array}
\end{displaymath}

This recovers the remarkable effect of the valley selectivity for circularly polarized photons. It is important to observe that although the excitonic character of the observed optical transitions at the fundamental gap is very pronounced~\cite{HeinzRMP}, this does not alter the selection rules derived from one-electron Bloch functions at the relevant edges, as the excitonic wavefunction satisfies those symmetries. 
%

%
\subsection{Symmetry character of core levels at \texorpdfstring{$\mathbf{K_+},  \mathbf{K_-}$}{Lg}}

As already mentioned in the Introduction, the core electron levels have very distinctive features when compared to valence states: one is their very localized character, so that, when described in terms of bands in momentum space, they have negligible dispersion and very large effective masses. In addition, the spin-orbit splitting of core levels (except the $s-$like ones) is much larger than for valence states and constitutes an energy scale exceeding the fundamental band gap and in most cases the typical bandwidth of valence levels (see a compilation of atomic binding energies of $Mo$ and $W$ in Appendix ~\ref{appendix:table}). Before dealing with the discussion of strongly spin-orbit-split $p-$ and $d-$core levels, let us briefly address the $s-$like core levels. This can in point of fact proceed very quickly, as the discussion parallels completely that of the conduction band bottom. From the point of view of the $C_{3h}$ group, the conduction band Wannier function $\Phi_{2,0}$ and the s-like function $\Phi_{0,0}$ have exactly the same symmetry properties of invariance under all group operations. Therefore we obtain the symmetry classification of $\Gamma_{10}, \Gamma_{12}$ for the two spin orientations at  $\mathbf{K_+}$ and $\Gamma_{9}, \Gamma_{11}$ at $\mathbf{K_-}$. This is enough to conclude that, in p-type samples, transitions to the valence band top shall display the valley selectivity of polarised photons. Transitions to the conduction band minima, corresponding to states with symmetry identical to the $s-$ core levels, are forbidden.

Going back to the strongly spin-orbit split $p-$ and $d-$ levels, it appears appropriate to modify the description of Bloch functions as given in Eq.~(5) or (12) by diagonalising at the very start the spin-orbit interactions and replacing the eigenfunctions of $\mathit{l}, m$ and $\sigma_{z}$ with those of $j$ and $j_{z}$. Given the very nearly dispersionless nature of the core bands, all the $2j+1$ values of $j_{z}$ are degenerate, to a very good approximation, throughout the Brillouin zone and constitute a basis for the $D_{j}^{\pm}$ representation of $O(3)$; the $\pm$ sign ambiguity depends on the parity of the spatial wavefunctions, e.g. odd for $p_{1/2}$ or  $p_{3/2}$ levels, even for $d_{3/2}$ or  $d_{5/2}$, etc. 

Thus the general form of a core wavefunction is
\begin{equation}
\Psi_{n,\mathit{l},j,j_{z} \mathbf{K_\pm}} = \sum_{j_{1},j_{2}} e^{i\mathbf{K_\pm}\cdot \mathbf{R}_{j_{1},j_{2}}} \Phi_{n,\mathit{l},j,j_{z}}(\mathbf{r-R}_{j_{1},j_{2}}).
\end{equation}
From the discussion of the valence levels, we learned that the Bloch factor in each addendum transforms like $\Gamma_{3}$ for $\mathbf{K_+}$ and like $\Gamma_{2}$ for $\mathbf{K_-}$. Therefore the complete Bloch function transforms like the product representation
\begin{equation} 
 \mathbf{K_+}:\;\;\;\;\Gamma_{3} \otimes D_{j}^{\pm}; \;\;\;\;\;   \;\;\;\ \mathbf{K_-}: \;\;\;\;\Gamma_{2} \otimes D_{j}^{\pm},
\end{equation}
with the \textit{odd} $D_{j}^{-}$ representation for $p $ core levels, and the \textit{even} $D_{j}^{+}$ for the $d-$like levels.

From tabulated results~\cite{KDWS-t58} we have the following reductions
\begin{equation}
D_{1/2}^{-} = \Gamma_{9} + \Gamma_{10},
\end{equation}
\begin{equation}
D_{3/2}^{-} = \Gamma_{9} + \Gamma_{10}+ \Gamma_{11} + \Gamma_{12},
\end{equation}
which allow to identify the reduction of the complete Bloch functions for $p_{1/2}$ and $p_{3/2}$ functions, i.e. for $L_{2}, L_{3}, M_{2}, M_{3}, N_{2}, N_{3}$,... core levels.
Starting with the $p_{1/2}$ levels at the $\mathbf{K_+}$ point, from Eq.~(21) they belong to the following representations
\begin{equation}
\Gamma_{3} \otimes \Gamma_{9} = \Gamma_{8};  \;\;\;\;\;  \Gamma_{3} \otimes \Gamma_{10} = \Gamma_{11}
\end{equation}
and at the $\mathbf{K_-}$ point, to the representations
\begin{equation}
\Gamma_{2} \otimes \Gamma_{9} = \Gamma_{12};  \;\;\;\;\;  \Gamma_{2} \otimes \Gamma_{10} = \Gamma_{7}.
\end{equation}

We are then in a position to derive the selection rules for optical transitions to the conduction band minima, using again Eq.~(13-16). The only allowed transitions are \\
at $\mathbf{K_+}$:
\begin{displaymath}
\begin{array}{lcr}
Initial &     Final &    Polarization \\
\Gamma_{8} & \Gamma_{12} &  + \\
\Gamma_{11} & \Gamma_{10} &  - 
\end{array}
\end{displaymath}
at $\mathbf{K_-}$ :
\begin{displaymath}
\begin{array}{lcr}
Initial &     Final &    Polarization \\

\Gamma_{12} & \Gamma_{9} &  +\\
\Gamma_{7} & \Gamma_{11} &  -
\\
\end{array}
\end{displaymath}

We therefore conclude that the valley selectivity effect for dipole transitions to the conduction band edges \textit{does not occur} at the $p_{1/2}$-like absorption edges $L_{2}, M_{2}, N_{2}$. This includes the cases of intrinsic or $n-$type samples, where the Fermi level is in the gap or in the conduction bands. For samples, where empty states near the top of the valence bands (i.e. in the $\Gamma_7$ and/or $\Gamma_8$ bands) occur, on the other hand, the selection rules for transitions from the $p_{1/2}$ core levels are
\\
at $\mathbf{K_+}$: 
\begin{displaymath}
\begin{array}{lcr}

Initial &     Final &    Polarization \\
\Gamma_{11} & \Gamma_{7} &  + 

\end{array}
\end{displaymath}
\\
at $\mathbf{K_-}$: 
\begin{displaymath}
\begin{array}{lcr}

Initial &     Final &    Polarization \\

\Gamma_{12} & \Gamma_{8} &  -
\end{array}
\end{displaymath}

In this case of $p-$type samples, therefore, the valley selectivity should be in principle observable at the $L_{2}, M_{2}, N_{2}$ absorption edges. Also, x-ray emission deriving from transitions from the full valence band states to core) holes should obey related selection rules (interchanging initial and final states and the sign of the polarisation). 

We now consider the $p_{3/2}$ levels. From Eq.~(22) one obtains that, at the $\mathbf{K_+}$ point they correspond to the following representations
\begin{widetext}
\begin{equation}
\Gamma_{3} \otimes \Gamma_{9} = \Gamma_{8};  \;\;\; \Gamma_{3} \otimes \Gamma_{10} = \Gamma_{11}; \;\;\; \Gamma_{3} \otimes \Gamma_{11} = \Gamma_{7};  \;\;\; \Gamma_{3} \otimes \Gamma_{12} = \Gamma_{9},
\end{equation}
and at $\mathbf{K_-}$:
\begin{equation}
\Gamma_{2} \otimes \Gamma_{9} = \Gamma_{12};  \;\;\;\ \Gamma_{2} \otimes \Gamma_{10} = \Gamma_{7}; \;\;\; \Gamma_{2} \otimes \Gamma_{11} = \Gamma_{10};  \;\;\;\ \Gamma_{2} \otimes \Gamma_{12} = \Gamma_{8}.
\end{equation}
\end{widetext}
Notice that $\Gamma_{7}$ and $\Gamma_{8}$ levels are present at both points, and we know from the discussion of the valence band transitions that these symmetries imply the valley selectivity of optical absorption to the conduction bands. It remains to investigate which transitions are allowed  from the $\Gamma_{9}$ and  $\Gamma_{11}$ levels to the  $\Gamma_{10}$ and  $\Gamma_{12}$ levels at $\mathbf{K_+}$, and from the $\Gamma_{10}$ and  $\Gamma_{12}$ levels $\Gamma_{9}$ and  $\Gamma_{11}$ levels at $\mathbf{K_-}$. The analysis based on Eq.~(14) gives the following results for allowed transitions \\
at $\mathbf{K_+}$:
\begin{displaymath}
\begin{array}{lcr}

Initial &     Final &    Polarization \\

\Gamma_{7} & \Gamma_{10} &  + \\

\Gamma_{8} & \Gamma_{12} &  + \\

\Gamma_{9} & \Gamma_{12} &  - \\

\Gamma_{11} & \Gamma_{10} &  -

\end{array}
\end{displaymath}
at $\mathbf{K_-}$:
\begin{displaymath}
\begin{array}{lcr}
Initial &     Final &    Polarization \\

\Gamma_{7} & \Gamma_{11} &  - \\

\Gamma_{8} & \Gamma_{9} &  - \\

\Gamma_{10} & \Gamma_{11} &  + \\

\Gamma_{12} & \Gamma_{9} &  + 
\end{array}
\end{displaymath}
Therefore, both valleys are accessible with both polarisations, and no valley selectivity holds for transitions to the conduction bands. Once again, considering $p-$type samples, (or x-ray emission spectroscopy) we obtain on the other hand, for $\Gamma_7$, $\Gamma_8$ final states\\
At $\mathbf{K_+}$: 
\begin{displaymath}
\begin{array}{lcr}

Initial &     Final &    Polarization \\
\Gamma_{11} & \Gamma_{7} &  + \\
\Gamma_{9} & \Gamma_{8} & +
\\
\end{array}
\end{displaymath}
At $\mathbf{K_-}$: 
\begin{displaymath}
\begin{array}{lcr}

Initial &     Final &    Polarization \\
\Gamma_{10} &  \Gamma_{7} &  -\\
\Gamma_{12} & \Gamma_{8} &  -
\end{array}
\end{displaymath}
Therefore one can expect valley selective absorption in $p-$type samples, or emission from valence band states to core holes at $L_{3}, M_{3}$ and $N_{3}$ edges.

Finally, let us now turn our attention to the $d-like$ core levels, that are split into $d_{3/2}$ and $d_{5/2}$, respectively corresponding to the representations $D_{3/2}^{+}$, $D_{5/2}^{+}$. From~\cite{KDWS-t58} the reduction to irreducible $C_{3h}$ representation gives
\begin{equation}
D_{3/2}^{+} = \Gamma_{7} + \Gamma_{8} + \Gamma_{11} + \Gamma_{12} 
\end{equation} 
\begin{equation}
D_{5/2}^{+} = \Gamma_{7} + \Gamma_{8} + \Gamma_{9} + \Gamma_{10} + \Gamma_{11} + \Gamma_{12} .
\end{equation} 

\begin{figure*}[t]
\includegraphics[width=17cm]{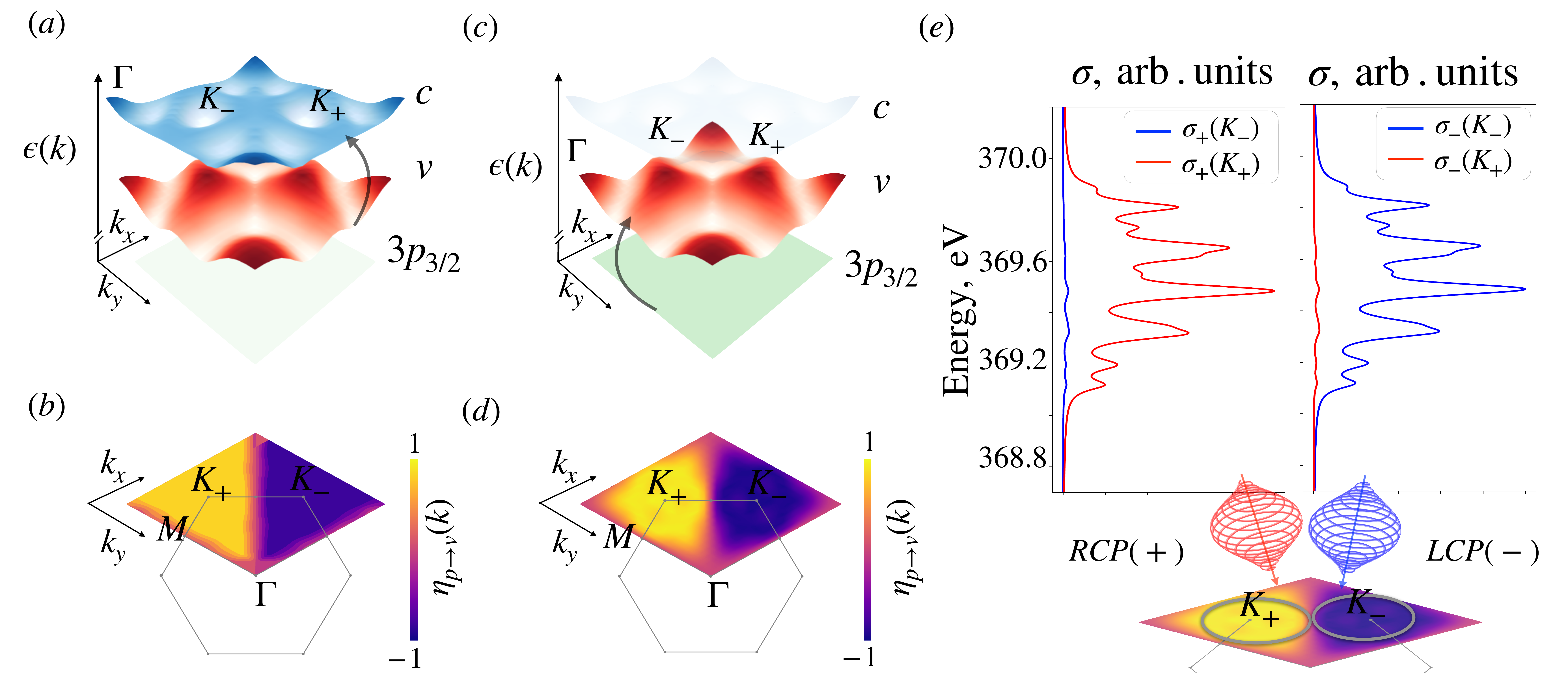}
\caption{Valley selectivity for transitions from the $Mo$ fourfold degenerate $3p_{3/2}$ level to the uppermost valence bands. (a),(c) -Sketchs of the valence to conduction band transition (a) and the core to valence band transition (c); (b),(d) - Computed normalized degree of optical polarization $\eta$ (see text) as a function of quasi-momentum \textbf{k} in the relevant portion of the Brillouin zone for the valence to conduction and core to valence band transitions, respectively;  (e) - Partial absorption cross-section ($\sigma$) obtained through integration over a circular area around $K_{\pm}$ points for left (-) and right (+) circular polarisation of the incoming x-ray photons (see also text and illustration of the absorption process in the bottom of (e)).}
\label{fig:figure3_core_to_valence}
\end{figure*}

We thus see that the sixfold degenerate $d_{5/2}$ state contains \textit{all} six additional irreducible representations of $C_{3h}$. They shall also all appear both at $\mathbf{K_+}$ ($\mathbf{K_-}$), because multiplication by $\Gamma_{2}$ ($\Gamma_{3}$) of the whole list of representations shall deliver a permutation of the same list. This tells us without further calculations that the valley selectivity \textit{does not occur} at $d_{5/2}$ edges, where all possible symmetries and transitions occur at both   $\mathbf{K_{\pm}}$ points. An explicit calculation shows that the selectivity is lost also at the $d_{3/2}$ edges. In fact the fourfold degeneracy of this state is split into four representations, namely 
\begin{widetext}
\noindent  at $\mathbf{K_+}$:
\begin{equation}
\Gamma_{3} \otimes \Gamma_{7} = \Gamma_{10};  \;\;\; \Gamma_{3} \otimes \Gamma_{8} = \Gamma_{12}; \;\;\; \Gamma_{3} \otimes \Gamma_{11} = \Gamma_{7};  \;\;\; \Gamma_{3} \otimes \Gamma_{12} = \Gamma_{9}
\end{equation}
and at $\mathbf{K_-}$:
\begin{equation}
\Gamma_{2} \otimes \Gamma_{7} = \Gamma_{11};  \;\;\;\ \Gamma_{2} \otimes \Gamma_{8} = \Gamma_{9}; \;\;\; \Gamma_{2} \otimes \Gamma_{11} = \Gamma_{10};  \;\;\;\ \Gamma_{2} \otimes \Gamma_{12} = \Gamma_{8};
\end{equation}
\end{widetext}

This combination of symmetries is not verifying valley selectivity. For example, we find at $\mathbf{K_+}$:

\begin{displaymath}
\begin{array}{lcr}
Initial &     Final &    Polarization \\

\Gamma_{9} & \Gamma_{12} &  - \\

\Gamma_{7} & \Gamma_{10} &  + 

\end{array}
\end{displaymath}

and similarly at $\mathbf{K_-}$:

\begin{displaymath}
\begin{array}{lcr}

Initial &     Final &    Polarization \\

\Gamma_{10} & \Gamma_{11} &  + \\

\Gamma_{8} & \Gamma_{9} &  -

\end{array}
\end{displaymath}

so that transitions with both polarisations are allowed in both valleys. It is also easily verified that no selectivity is present in $p-$type samples either. In the following section we will use first-principle calculations to confirm all the selection rules obtained for high symmetry points, and extend them through large regions of the  Brillouin zone. 
\section{Numerical calculations of transition rates}

\label{sec-3} 
In order to confirm the findings of the group-theoretical analysis of the selection rules for the $\mathbf{K_+}$ and  $\mathbf{K_-}$ band edges, DFT \textit{ab initio} calculations were performed throughout the Brillouin zone for the prototypical $MoS_2$ MTMD, using the all-electron \textit{exciting} code~\cite{Gulans2014} with linearized augmented plane-wave and local orbitals (LAPW+lo) basis set~\cite{Andersen1975}. In this way, the rapid oscillation of the localized wave functions around the nuclear position is handled in the local atomic orbital basis within the sphere of the "muffin-tin" radii and the rest of the solid (interstitial region) in the plane-wave basis.  Due to the dominance of the relativistic effects for the core states, we obtained core-state wave functions by solving the set of coupled Dirac equations in a spherically symmetric potential as implemented in \textit{exciting}~\cite{Gulans2014}. In contrast, the conduction and valence states were obtained in the scalar-relativistic approximation.  The latter also implies that the small components of the 4-spinor wave function were neglected when calculating the transition matrix elements. 
We employ the generalized gradient approximation (GGA) PBE~\cite{PBE1996} for exchange-correlation functional, a momentum space grid of $21\times 21 \times 1$ points, and a muffin-tin radius of 2 a.u. for both species.

 

\begin{figure*}[t]
\includegraphics[width=17cm]{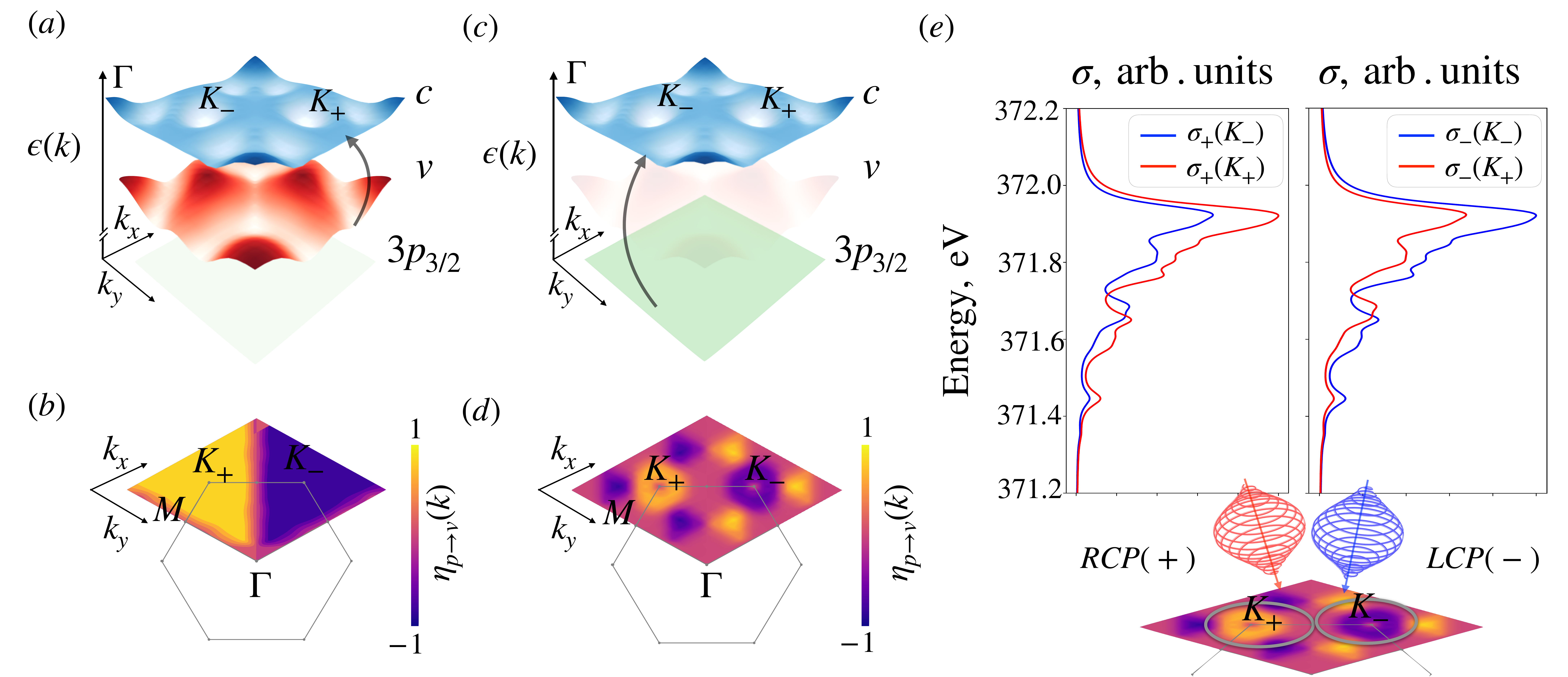}
\caption{Valley selectivity for transitions from the $Mo$ fourfold degenerate $3p_{3/2}$ level to the lowest conduction band. (a),(c) -Sketch of the valence to conduction band transition (a) and the core to conduction band transition (c); (b),(d) - Computed normalized degree of optical polarization $\eta$ (see text) as a function of quasi-momentum \textbf{k} in the relevant portion of the Brillouin zone for the valence to conduction and core to conduction band transitions, respectively; (e) - Partial absorption cross-section ($\sigma$) obtained through the integration over circular area around $\mathbf{K}_{\pm}$ points for the left (-) and right (+) circular polarisation of the incoming x-ray photons (see also text and illustration of the absorption process in the bottom of (e)).}
\label{fig:figure4_core_to_conduction}
\end{figure*}

 
 The selected core level is the $Mo$ $3p_{3/2}$, or $M_3$ level, which (see Table I) is located about $390 \, eV$ below the Fermi energy. The quantity that is computed and plotted is the normalized "degree of optical polarization" $\eta$ as defined in ref.~\cite{Cao2012}, for the transitions from the four degenerate $p_{3/2}$ levels to the band $b$
\begin{equation}
\eta_{p_{3/2}-b} (\mathbf{k}) = \sum_{j=-3/2}^{3/2}\frac{|D_{+}^{jb} (\mathbf{k})|^{2} - |D_{-}^{jb} (\mathbf{k})|^{2}} {|D_{+}^{jb} (\mathbf{k})|^{2} + |D_{-}^{jb} (\mathbf{k})|^{2}},
\end{equation}
where
\begin{equation}
D_{\pm}^{jb}  (\mathbf{k}) = [D_{x}^{jb}  (\mathbf{k}) \pm iD_{y}^{jb}  (\mathbf{k})] 
\end{equation}
and $D$ is the appropriate component of the momentum matrix element from the core level $j$ to the valence or conduction band edge $b$. The results for transitions to the valence and to the conduction band are shown in Figures 3 and 4 respectively. Furthermore, to show the energy dependence of the corresponding transition, we calculated the absorption spectra ($\sigma$) in the independent particle limit in vicinity of $K_{\pm}$ points
\begin{equation}
\sigma_{\pm}(\omega)= 
\sum_{j,b} \int\displaylimits_{\Omega(k_{max})} \hspace*{-1em} d {\textbf{k}}  |D^{jb}_{\pm}({\bf{k}})|^2  \delta(\omega - (E_b ({\textbf{k}})-{E_j} ({\textbf{k}})) .
\end{equation}
Here the integration area $\Omega(k_{max})$ is defined by the parameter $k_{max}$ as $|\bf{k}-\bf{K_{\pm}}|<\mathit{ k_{max}}$. The latter is chosen to be the half of the distance between the $\bf{K}_+$ and $\bf{K}_{-}$ points. The transition matrix elements $M_{jb}(\bf{k})$ between core and valence (conduction) states were calculated in the dipole approximation and $E_j$, $E_b$ are the energies of the corresponding levels. A small Lorentzian broadening was attributed to the resulting spectra for visual purposes. Such partial absorption cross-sections are equal for left and right circular polarization of the incoming photon if there is no valley selectivity at a given energy (Figure~\ref{fig:figure4_core_to_conduction}). The dominant contribution from the corresponding valley ($\mathbf{K}_+$ for right circular polarization and $\bf{K}_{-}$ for the left), on the other hand, highlights strong selectivity (See Figure~\ref{fig:figure3_core_to_valence}). 
%

\begin{figure}[h!]
\includegraphics[width=8cm]{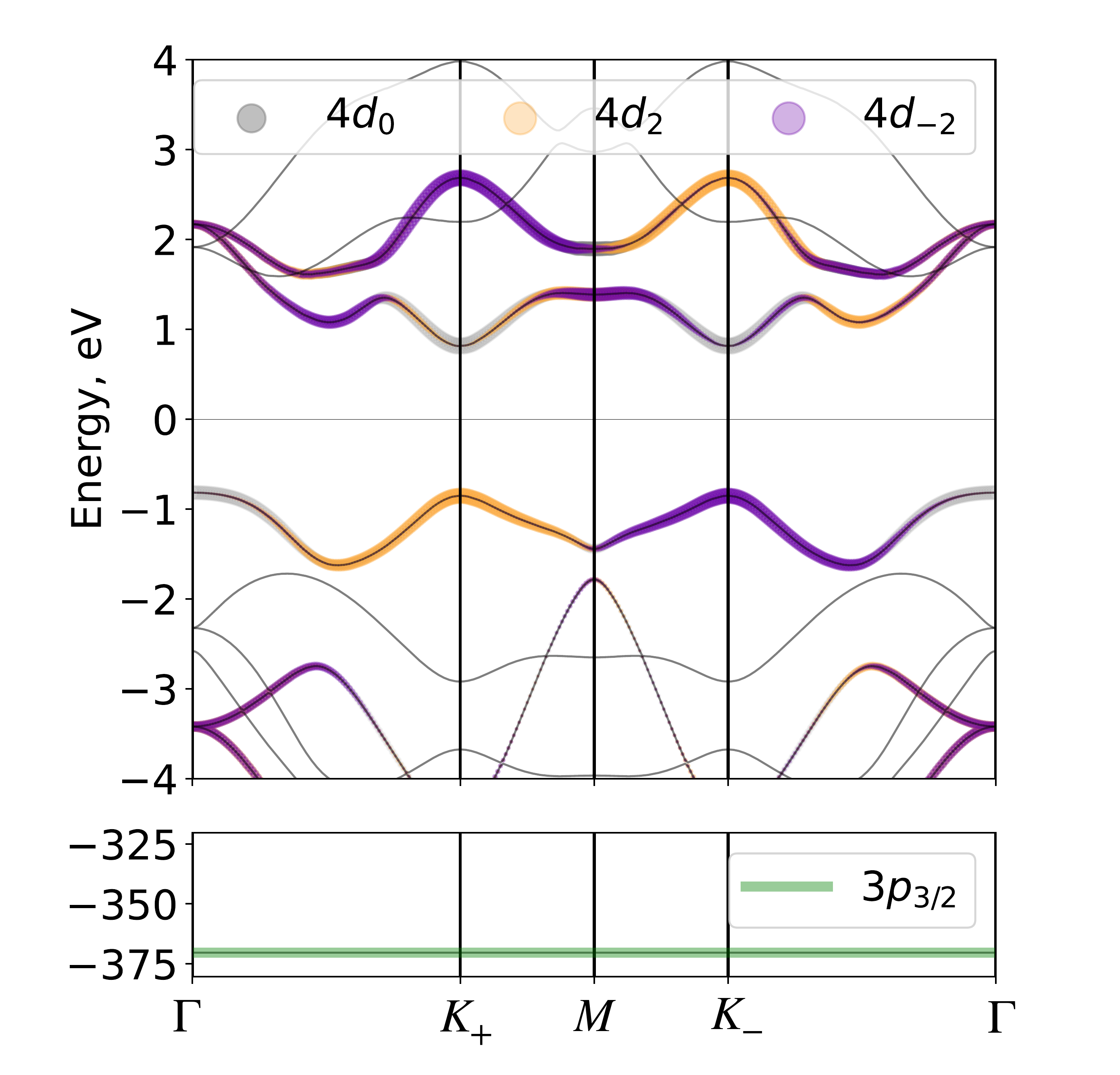}
\caption{Band structure of $MoS_2$ monolayer,  where we highlight the  projection of the Kohn-Sham wavefunction onto the $4d_m \  m=\{0,\pm 2\}$ and $3p_{j=3/2}$ orbitals of the molybdenum atom.  }
\label{fig:figure45}
\end{figure}

The calculations, besides confirming the opposite value of $\eta$ at the $\mathbf{K_+}$, $\mathbf{K_- }$ points for valence band transitions, also reveal that the strong polarisation character persists in very large portions of the Brillouin zone, as one moves away from these points; a fact that would not be easily predictable by group theory, but that is in complete analogy to the findings of Cao \textit{et al.}~\cite{Cao2012} for transitions at the fundamental gap. This is an aspect that is important for soft-x-ray spectroscopy, because it implies that the valley selectivity is not going to be washed out by the relatively larger bandwidth of x-ray pulses. The DFT calculations also underline very different behaviour of transitions to the conduction bands, where, besides confirming the zero selectivity at the $\mathbf{K_\pm}$ points, they reveal a very different distribution of  $\eta$ in the two regions surrounding the points. In order to study the role of the orbital character in the selection rules for the whole Brillouin we performed a projection of the Kohn-Sham wavefunctions onto the basis of the local orbitals of the molybdenum atom and results are presented in Figure~\ref{fig:figure45}.
Qualitatively, instead of almost uniform distribution, which we observed for the core to valence band transition, the $\eta$ distribution for the core to conduction band transitions shows two types of sub valleys with the opposite signs. The first type comprises the valleys around the $K_{\pm}$ point, and the second the small triangular valleys in between $K_{\pm}$ and $\Gamma$. The latter corresponds to the local minimum of the conduction band (Figure~\ref{fig:figure45}). The opposite selectivity of the two sub valleys is related to the $d$ orbital character of the conduction bands where $d_{2}$ and $d_{-2}$ interchange each other around local and global minima. From Figure~\ref{fig:figure45} we could also see that for the conduction band in the close vicinity of the $K_{\pm}$ the $d_{0}$ type of orbitals have the most substantial contribution and correlate with the absence of dichroism predicted from group theory. Furthermore, for the valence band, orbital characters $d_{\pm 2}$ are dominant in $K_{\pm}$ valleys respectively, which expand almost up to $\Gamma$ point and lead to the more isotropic distribution of $\eta$. In line with the conclusion of the previous section our \textit{ab initio} calculations highlight that the symmetry of the local orbitals involved in the transition lies at the very origin of the valley selectivity.  
\section{Soft-x-ray core level spectroscopy: applications to "valleytronics"}
\label{4-section}

\begin{figure*}
\includegraphics[width=17cm]{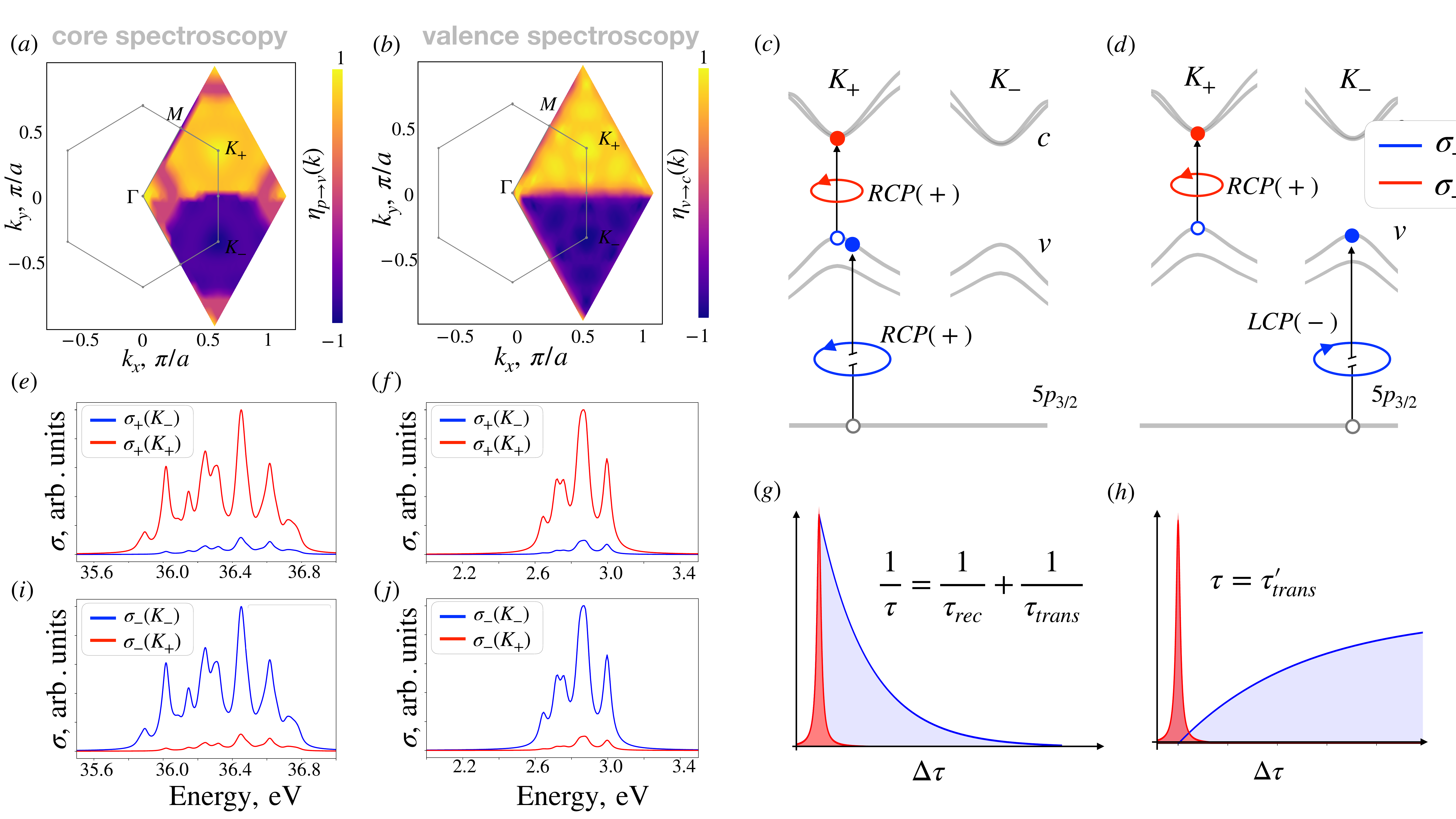}
\caption{ Valley selectivity of $WS_2$ and schematic of pump-probe experiment. (a), (b) - Normalized degree of optical polarisation as a function of quasi-momentum for transitions from $W$ $ 5p_{3/2}$ levels to the uppermost valence band (a), and from the uppermost valence band to the lowest conduction band (b). (f), (j) ((e), (i)) - Partial absorption cross-sections calculated in vicinity of ${\bf{K}_\pm}$ points for the valence to conduction (core to valence) band transitions. (c), (d) - Schematic representation of the core to valence and valence to conduction bands transitions for the optical and x-ray photons with the same (c), and opposite (d), circular polarization. The vertical red arrows denote transitions from IR/visible laser pump pulses, the blue arrows transitions from core levels. (g), (h) - Schematic figures of the pump pulse (in red) and the transient probe absorption (in blue) as a function of pump-probe delay $\Delta \tau$ for the configuration of polarization presented in (c) and (d), respectively. See text for the definition of the various characteristic times $\tau$. }
\label{fig:figure5_experiment}
\end{figure*}

The main result of the previous sections is the valley selectivity of transitions from the metal $s$ and $p$ core levels to the valence bands. In order to have empty states near the valence band top, the available options are gated devices, typically with a MTDM sandwiched between hBN layers with electrodes at the bottom and the top~\cite{MakShan}, or optical pumping, as in ref.~\cite{Attar, Chang, Britz} and possibly $p-$type doping~\cite{Tang, Liang}.
 If we postpone for the time being the discussion on the feasibility of XUV and soft-x-ray transmission measurements in atomically thin samples, novel diagnostic tools for valleytronics can be envisaged. For example, taking inspiration from the pump-probe transient absorption experiments reported by Chang \textit{et al.}~\cite{Chang} at the $W$ $5p_{3/2}$ edge of multilayer $WS_2$ samples, consider the pump-probe experiments for the monolayer case schematically displayed in Figure~\ref{fig:figure5_experiment}(c-d). Here we propose to use circularly polarised pump and probe pulses; we argue that a circularly polarised pump pulse should induce a pronounced \textit{circular dichroism} in the pump absorption. If the pump and the probe polarisations are the same (Figure~\ref{fig:figure5_experiment}(c)), the transient absorption of the probe should be observed at very small delay, because the transitions from the core levels access the valley in which a hole population was created by the pump. As the delay of the two pulses increases the transient absorption should be attenuated, with characteristic time $\tau$, given by
\begin{equation}
1/\tau = 1/\tau_{rec} +1/\tau_{trans},
\end{equation}
because the hole population is reduced by recombination (rec) and by transfer(trans) of holes to other valleys. If the probe polarisation is inverted (Figure~\ref{fig:figure5_experiment}(d)), the probe can now promote core electrons only to the valley opposite to the one populated with holes by the pump: the absorption signal is therefore much smaller than in the previous case at short delay times, and can possibly build up in time, if sufficient transfer to the opposite valley takes place, with characteristic time
\begin{equation}
1/\tau = 1/\tau_{trans}',
\end{equation}
(where the prime acknowledges that only transfers to the opposite valley are included, and not e.g. to the $\Gamma$ valley). At any delay time, therefore, the absorption of the pump pulse should display circular dichroism and, according to Eq.~(33 - 34), should deliver information on the characteristic decay times of the valence valley polarisation.

To consider specifics related to $WS_2$ monolayer, we performed calculations of the normalized polarisation degree for transitions first from the core levels ($5p_{3/2}$ of W) to the uppermost of the valence bands (Figure~\ref{fig:figure5_experiment}(a)) and then from the uppermost of the valence bands to the lowest of the conduction bands (Figure~\ref{fig:figure5_experiment}(b)). As a result, both core and optical transitions show pronounced valley selectivity with the same sign in $\bf{K}_{+}$ and $\bf{K}_{-}$ valleys. The latter means that if circular polarisation of the optical and x-ray photons are the same, we can probe information about the single valley; and if on the other hand they are different, contributions from cross valleys processes will be dominant. In contrast to the previous section, calculations for $\mathrm{WS_2}$ include relativistic effects for core and valence electrons and consequently reproduce spin-orbit splitting of $\sim 8\ eV$ for 5$p$ states of tungsten, and $\sim 0.3$ eV for the valence $d$ states. The results are presented in Figure~\ref{fig:figure5_experiment} and agrees with the group theoretical predictions. It is also worth noticing that the two valleys are extended through the most part of the Brillouin zone and manifest strong difference in partial absorption. In Figure~\ref{fig:figure5_experiment}(a)(e)(i) we are showing only characteristic contributions from those $5p_{3/2}$ fourfold degenerate levels that participate in non-zero transitions at the $\mathbf{K_{\pm}}$ points, contribution from the rest of 5$p_{3/2}$ levels favorably support the conclusion in the extended valley region but lead to the divergence of the polarization degree due to the zero denominator. As expected from the group theoretical analysis the core to the lowest conduction band transitions are not valley selective at $\bf{K}_{\pm}$, but have certain selectivity for the small parts of the Brillouin zone (see also Appendix C). Following the approach stated in the previous section we obtained partial absorption cross-sections for the both types of transitions which confirms strong dichroism through the entire energy range of the highest (lowest) valence (conduction) bands. %
%

%
It is generally presumed that intervalley transfer occurs for excitons, via exciton-exciton or exciton-phonon scattering, with much higher probability than for single carriers (electrons or holes) as the latter case implies either a spin flip or a sizeable energy difference~\cite{HeinzRMP}. Nonetheless, it is worthwhile to notice that the core electron circular polarisation transitions are strictly a probe of the hole population, unlike e.g. photoluminescence. In special cases, such as the presence of non-radiative recombination or trapping centers for either carrier, the distinction may become relevant. 

 The actual feasibility of core level absorption spectroscopy on single layer samples deserves discussion, in particular in the pump-probe mode as discussed above. With respect to the results reported in ref.~\cite{Chang}, there is on the one hand the reduction from several tens of layers to one, and the requirement of circularly polarised photons on the other. To the former challenge, the possible strategies could involve the growth of samples in which single layers are stacked with a $hBN$ spacer in between; and also the selection of a more intense source. By replacing the HHG source used in ref.~\cite{Chang} with a free-electron laser such as the FERMI facility~\cite{Fermi}, the intensity increases by about $2$ to $3$ orders of magnitude; and one can ripe the additional benefit of satisfying the requirement of circularly polarised photons.
 
 Another technique to explore the intervalley relaxation of carriers is time-resolved Kerr rotation (TRKR) \cite{Dey,Ersfeld} which has been used in samples where tuning the Fermi level by the gate potential produces, independent of optical pumping, an equilibrium population of electrons or one of holes. In presence of these so-called \textit{resident} carriers, optical pumping with circularly polarized photons near the band gap produces, in one of the valleys, a rapidly decaying population of charged excitons, or trions, which results in an upset of the balance in the population of the two valleys. The imbalance corresponds to a nonzero value of the  orbital angular momentum. The exact role of the trions decay and the role of non radiative traps due to impurities is not completely clear \cite{Ersfeld}. In any event, the measurement of the decay time of the Kerr rotation allows access to the decay times of the difference in valley population of electrons in n-type samples and of holes in p-type samples. The latter case is especially interesting as decay times as long as  several $\mu s$ are observed. In these samples, generation of the holes imbalance between the valleys could be produced by transitions from the core levels with circularly polarized x-ray photons to the valence bands, without intermediate stages in which excitons or trions are present. The core hole would be very rapidly filled by Auger processes or by x-ray fluorescence from electron levels from the whole valence bands; such core hole annihilation should not involve any valley selectivity and should therefore not influence the Kerr rotation signal. This would therefore simplify the understanding of the imbalance relaxation paths and of the corresponding time scales.

 \section{Conclusions}
 In summary the conclusions of the present analysis of the selection rules for transitions in  core level spectroscopy in MTMD are that:
 
 - The valley selectivity of circular polarisation transitions to the conduction band minima is not to be expected for $s-$, $p-$ or $d-$like metal core levels, as shown by group-theoretical arguments and by \textit{ab initio} calculations, in contrast to the well-known top of the valence to conduction valleytronic selectivity.
 
 -  Valley selectivity is on the other hand expected for transitions from $s-$ and $p-$like core levels (but not for $d-$like levels)  to valence states (in contrast to the well known top of the valence-to-conduction selectivity, valleytronics), and conversely for x-ray emission.  This conclusion assumes of course that the valence bands are not completely full, as can be obtained in a static way by electrostatic biasing, or in a static or time-dependent way by optical pumping. 
 
 - We argued that these selection rules open the way to a new tool for the study and time resolved monitoring of valleytronics in 2D-TMD, and a specific example was discussed.

The potential applications, however, are not limited to the idea of using an x-ray pulse as a probe. In fact, another possibility to profit from derived selection rules is to use the core-to-valence transition under circularly polarized light to create an imbalanced population in the two valleys for the initially hole-doped TMD sample. That would allow more direct access to dynamics of the hole/electron carriers without creating multiparticle excitations (e.g. excitonis,trions), which accompany transitions in the optical region.
 
 Before further considering possible experimental implications of the results derived here for core-level transitions, the following points should be taken into account:
 
 1. Feasibility of specific experiments, depending on the actual cross sections, considering the extremely diluted character of monolayer samples.
 
 2. Excitonic effects. The valence to conduction spectroscopy is strongly affected by exciton effects~\cite{HeinzRMP}, as expected from considerations on screening of the electron-hole interaction in low-dimensional systems. The same arguments should apply to electron-core hole interactions, and core excitons can be expected to modify the transition energies and intensities, but not necessarily the symmetry considerations discussed here. In general if a pair of bands are connected by an allowed dipole transition, so is the corresponding ground exciton state.
 
 3. The just mentioned dipole approximation, i.e. the assumption that the photon momentum is negligible in comparison to the size of the Brillouin zone. This is reasonable for the shallowest core levels, like the M edges of $Mo$ or the N-edges of $W$, but certainly not for the deeper ones. 
 
 4. The lifetime broadening of core levels (especially the deeper ones) is a limitation of the achievable energy resolution. In some cases it can prevent a clear distinction of transitions to the $\mathbf{K_{\pm}}$ valleys from nearby secondary band extrema, requiring a specific discussion of each compound.

 Finally, we considered a material with close structural symmetry but without strong spin-orbit coupling such as hexagonal Boron Nitride (hBN). We confirmed valley selectivity for excitations in the optical region and derived corresponding selection rules. However, there are  very few dipole allowed transitions from core levels to the lowest  conduction and  valence band and in total they do not show the valley selectivity. 
 \begin{acknowledgments}
 We thank Kin-Fai Mak , Jie Shan and Claudio Masciovecchio for interesting discussions. This work was supported by the Cluster of Excellence ‘Advanced Imaging of Matter' (AIM), Grupos Consolidados (IT1249-19) and SFB925 "Light induced dynamics and control of correlated quantum systems. The Flatiron Institute is a division of the Simons Foundation. A.G. thank Dong Eon Kim and was supported by a National Research Foundation of Korea grant funded by the Ministry of Science and ICT No.~2016K1A4A4A01922028.
 \end{acknowledgments}

\appendix

\section{Valley selectivity without spin-orbit coupling: \texorpdfstring{$hBN$}{Lg}} 
\label{SM-hBN}
The purpose of this Appendix is to briefly consider the valley-selective effects in monolayer hexagonal boron nitride, $hBN$, with a simple extension of the discussion given in Section 2A. In fact, the lattice symmetry for a monolayer of $hBN$ is identical to that of 2D-TMD, with $D_{3h}$ as point group and $C_{3h}$ as the wavevector group at the $\mathbf{K_{\pm}}$ points~\cite{Doni}. According to the prevailing results of DFT theoretical work, the valence band top of monolayer $hBN$ is located at the  $\mathbf{K_{\pm}}$ points; whether the conduction band minimum is at $\Gamma$ or at $\mathbf{K_{\pm}}$ is a very close call~\cite{Ciraci, Ciraci2, Huang, Kim} and the computed gap is extremely sensitive to the chosen functional and to details of the numerical procedure~\cite{Nieminen}. Nonetheless, even if the lowest gap is indirect, optical properties are very likely dominated by the larger oscillator strength of the direct transition, and experimental evidence~\cite{Eaves} points in this direction.
\begin{figure*}
\includegraphics[width=15cm]{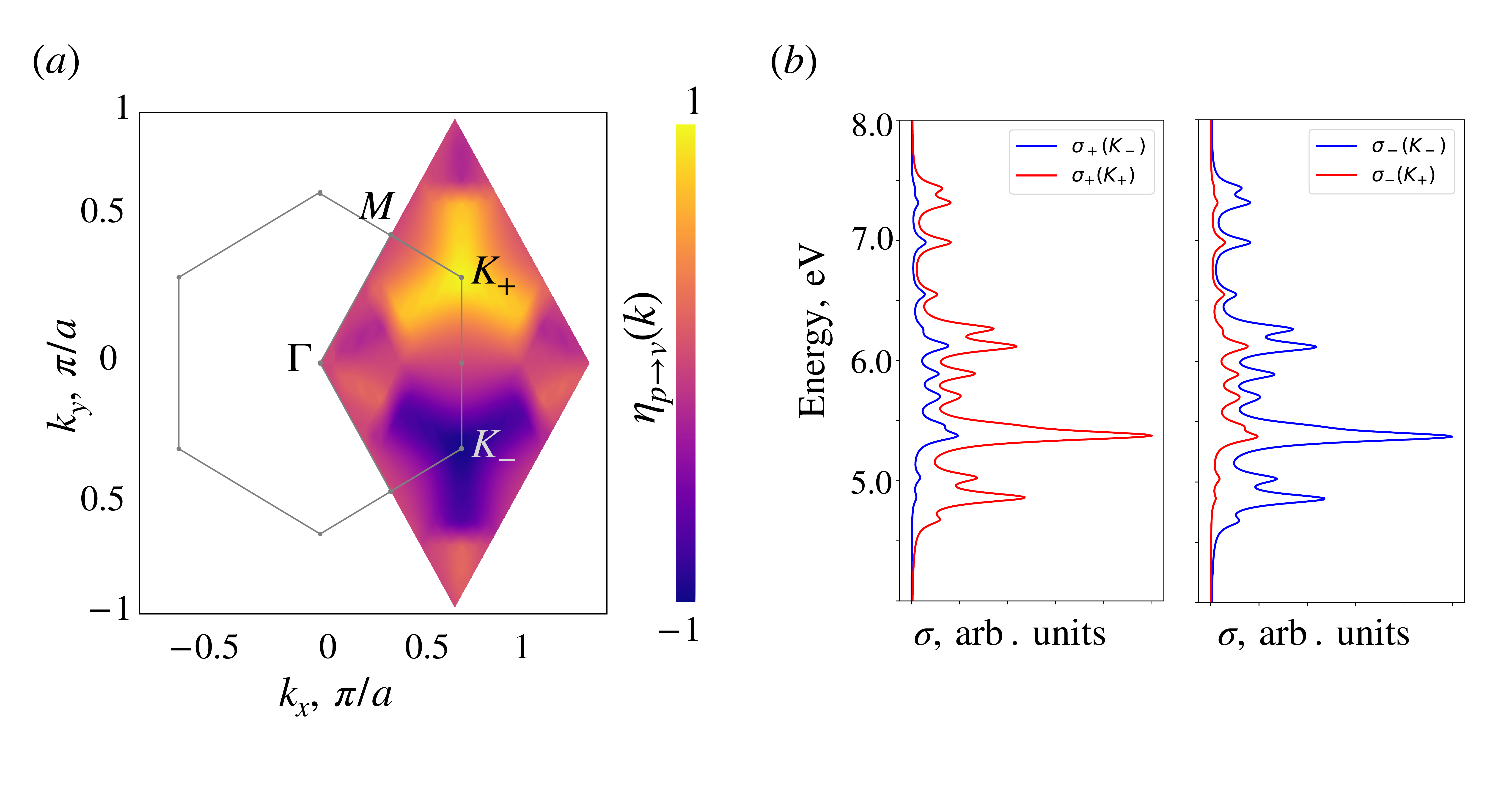}
\caption{(a) Computed normalized degree of optical polarization $\eta$ (see text) for transitions from the $hBN$ highest valence band the lowest conduction band, as a function of quasi-momentum \textbf{k} in the relevant portion of the Brillouin zone. (b) Partial absorption cross-section calculated for the circular area around $\bf{K_{\pm}}$ points for left (-) and right (+) circular polarization  of the incoming photons (see also Section 3). }
\label{fig:SM-optical-hBN}
\end{figure*}
\begin{figure*}
\includegraphics[width=15cm]{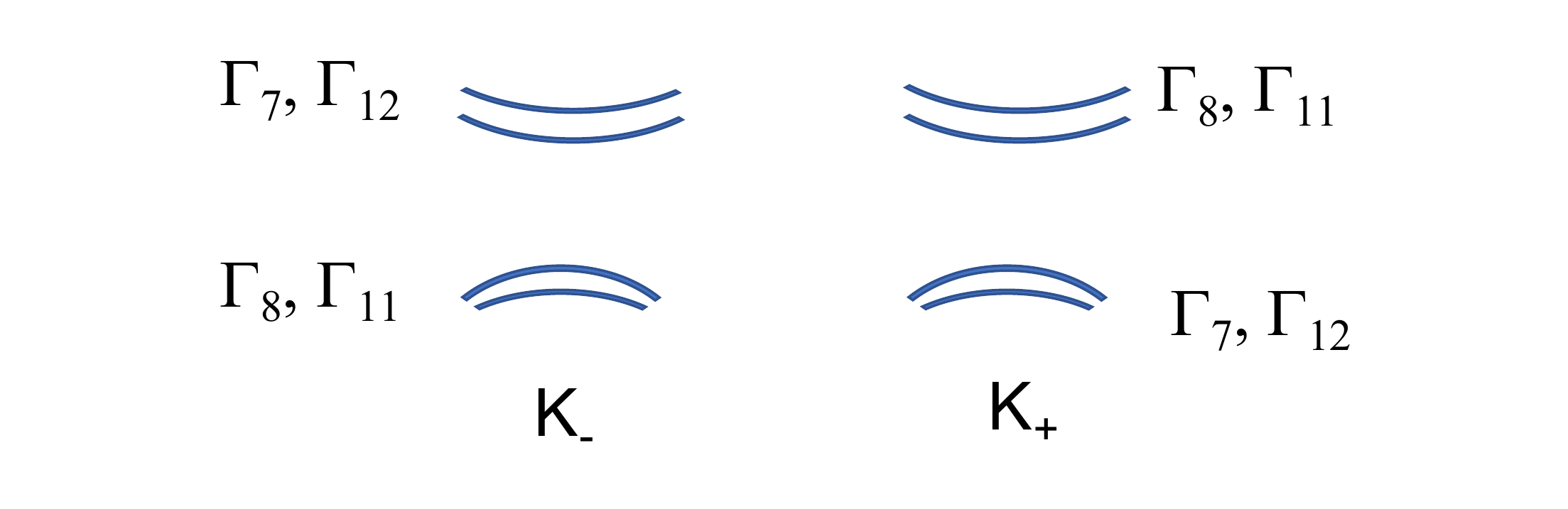}
\caption{Schematics of the valence and conduction band edges and their symmetry labels for $hBN$ at the $\mathbf{K}_{\pm}$ points.}
\label{fig:SM-hBN-sketch}
\end{figure*}

The group theoretical analysis can follow the pattern of Eq (5) and following of Section 2A, the only difference being the orbital character of the Wannier-like functions for $hBN$. According to the band structure calculations, see for example ref.~\cite{Ciraci}, the dominant character of the conduction band minima is derived from the $B$ $2p_z$ states, whereas the valence band top is ascribed to $2p_z$ orbitals of the $N$ atoms. Given the low atomic number of these elements, spin-orbit interactions are negligible. Starting with the conduction band minimum at $\mathbf{K_+}$, we observe that the $2p_z$ orbitals are invariant for all proper rotations about the $z$ axis ($m=0$), but are odd under all operations reversing the z axis directions, i.e. $\sigma_z$, $S_{3}^{\pm}$. A look at the character table of  $C_{3h}$ associates this behavior to $\Gamma_4$ representation. The symmetry character of the exponential Bloch factor for the metal sites is identical to the MTMD case and  belongs to the $\Gamma_3$ irreducible representation. Finally, the spin wavefunction belongs to the reduction of the $D_{1/2}^{+}$ representation that gives $\Gamma_{7} +\Gamma_{8}$. Putting all together, the conduction band minimum at $\mathbf{K_+}$ for the two spin directions is associated to  
\begin{equation}
\Gamma_{3} \otimes \Gamma_{4} \otimes \Gamma_{7} = \Gamma_{8}; \;\;\; \Gamma_{3} \otimes \Gamma_{4} \otimes \Gamma_{8} = \Gamma_{11}.
\end{equation}

Consider now the top valence bands, corresponding to similar $2p_z$ orbitals of $N$, centred however on the "empty" vertices of the lattice sketched in Figure 2. These sites can be indexed in analogy to Eq.~(6) as
\begin{equation} 
\mathbf{R}_{j_{1},j_{2}}^{N} = j_{1} \mathbf{a_1} + j_{2} \mathbf{a_2} + \mathbf{\delta_3} +\mathbf{\delta_2} .
\end{equation}
Note that with this indexing, since $\mathbf{K_+} \cdot \mathbf{\delta_2} = 0$, both  $j_{1}, j_{2}$ -indexed anion and cation sites belong to the same wavefront and have a common phase
\begin{equation} 
exp[i \mathbf{K_+}\cdot  \mathbf{R}_{j_{1},j_{2}}^{B,N} ]= exp[i(2j_{1} +j_{2} -1)\frac{2\pi}{3}].
\end{equation}

Repeating the procedure outlined in Appendix B, and noting that $ C_{3}^{-}  \mathbf{\delta_2}=\mathbf{\delta_3}$ we find that after a $C_{3}^{+}$ rotation, the phase factor of the $N$ sites becomes
\begin{equation} 
exp[i(-j_{1}-2 j_{2})\frac{2\pi}{3}]  \equiv  exp[i(2j_{1} + j_{2})\frac{2\pi}{3}],
\end{equation}
and, in general, $C_{3}^{\pm}, S_{3}^{\pm}$ produce a factor $e^{\pm(i2\pi/3)}$, which allows us to associate the Bloch factor at the $N$ sites to the $\Gamma_2$ representation. 

The valence bands analogous of Eq.~(A1) becomes then
\begin{equation}
\Gamma_{2} \otimes \Gamma_{4} \otimes \Gamma_{7} = \Gamma_{12}; \;\;\; \Gamma_{2} \otimes \Gamma_{4} \otimes \Gamma_{8} = \Gamma_{7}.
\end{equation}

It is readily seen that if we now consider the same bands at $\mathbf{K_-}$ we must interchange $\Gamma_{2}$ and $\Gamma_{3}$ in Eq.~(A1) and (A5), obtaining the band schematic diagram in Figure~\ref{fig:SM-hBN-sketch}. The usual procedure allows us to derive the selection rules for circularly polarized light
At $\mathbf{K_+}$:
\begin{displaymath}
\begin{array}{lcr}

Initial &     Final &    Polarization \\

\Gamma_{12} & \Gamma_{8} &  - \\

\Gamma_{7} & \Gamma_{11} &  -

\end{array}
\end{displaymath}

At $\mathbf{K_-}$:

\begin{displaymath}
\begin{array}{lcr}

Initial &     Final &    Polarization \\
\Gamma_{11} & \Gamma_{7} &  + \\

\Gamma_{8} & \Gamma_{12} &  +

\end{array}
\end{displaymath}

showing as valley selectivity effect at the $\mathbf{K}_{\pm}$ valleys of $hBN$. Our numerical calculations (Figure~\ref{fig:SM-optical-hBN}) confirms such analysis, however they also suggest that the valley are smaller then in case of $MTMD$ and elongate towards $\Gamma$ point. Moreover, we considered transitions from  $1s$ levels of Boron and Nitrogen to the lowest conduction and highest valence band respectively. Although some regions close to $\Gamma$ show some degree of dichroism, the direct transition from the core-levels to the $\bf{K_\pm}$ points with circularly polarized photons are forbidden for both species, and that prevents direct probing of the valleys. 


\section{}
 
 For completeness, we enumerate the steps leading to the conclusion that the space part of the Bloch function (i.e. the part that does not include spin variables) at $\mathbf{K_+}$ transforms as the $\Gamma_3$ representation of $C_{3h}$. The first observation is that the $\sigma_h$  leaves all factors unchanged, because atomic position vectors lie in the plane, $\Psi_{2,0} \sim (3z^{2} - r^2)$ is even and the spin is a pseudovector. Consider now the counterclockwise rotation by $2\pi /3, C_{3}^+$. To determine its effect on the spatial part of the wavefunction, i.e. on
\begin{equation}
\sum_{j_{1},j_{2}} e^{i\mathbf{K_+}\cdot \mathbf{R}_{j_{1},j_{2}}} \Phi_{2,0}(\mathbf{r-R}_{j_{1},j_{2}}),
 \end{equation} 
following Eq.~(8) we have to evaluate
\begin{equation}
 \sum_{j_{1},j_{2}} e^{i\mathbf{K_+}\cdot \mathbf{R}_{j_{1},j_{2}}} \Phi_{2,0}(C_{3}^{-}(\mathbf{r})-\mathbf{R}_{j_{1},j_{2}}).
 \end{equation} 
 This can be done by noticing that the sum is unchanged if each $\mathbf{R}_{j_{1},j_{2}}$ is replaced by $C_{3}^{-} (\mathbf{R}_{j_{1},j_{2}})$, i.e.
\begin{widetext}
 \begin{equation}
\sum_{j_{1},j_{2}} e^{i\mathbf{K_+}\cdot \mathbf{R}_{j_{1},j_{2}}} \Phi_{2,0}(C_{3}^{-}(\mathbf{r})-\mathbf{R}_{j_{1},j_{2}}) = \sum_{j_{1},j_{2}} e^{i\mathbf{K_+}\cdot C_{3}^{-}(\mathbf{R}_{j_{1},j_{2}})} \Phi_{2,0}(C_{3}^{-}(\mathbf{r}-\mathbf{R}_{j_{1},j_{2}})).
 \end{equation} 
\end{widetext}
This is because, when $j_{1}, j_{2}$ run over all integers, both sums on the right and on the left hand side run once over each and every metal site. 
To make further progress, we determine the lines of constant phase of the unrotated Bloch wave (the Bloch wavefronts, shown as dotted lines in Figure 2, the integer near the top of each line being $(2j_{1} + j_{2}\: mod \: 3)$); the phase factor being given by
\begin{equation}
e^{i \mathbf{K_+}\cdot (j_{1}\mathbf{a_1} + j_{2} \mathbf{a_2} + \mathbf{\delta_3}) }= e^{i(2j_{1} +j_{2} -1)\frac{2\pi}{3}}.
\end{equation}
To see the effect of the $C_{3}^-$ rotation, we notice that its effect is to transport the atom located at $j_{1} \mathbf{a_1} + j_{2} \mathbf{a_2} + \mathbf{\delta_3}$ to the new location $-(j_{1}+j_{2}) \mathbf{a_1}+j_{1} \mathbf{a_2} +\mathbf{\delta_1}$. This is readily seen by inspection of Figure 2, because:
\begin{equation}
C_{3}^{-}\mathbf{a_1} = \mathbf{a_{2}-a_{1}}; \;\;\;\; C_{3}^{-}\mathbf{a_2} =  -\mathbf{a_1}; \;\;\;\;\; C_{3}^{-}  \mathbf{\delta_3}=\mathbf{\delta_1}.
\end{equation}
The phase factor associated to the  $j_{1}, j_{2}$ term after the rotation becomes
\begin{equation}
e^{i(-j_{1}-2 j_{2} +1)\frac{2\pi}{3}} \equiv  e^{i(2j_{1} + j_{2}+1)\frac{2\pi}{3}},
\end{equation}
where the r.h.s. results from inserting factors  $e^{i3j_{i} \frac{2\pi}{3}} = 1, i=1,2$; comparing to Eq.(12) it is immediately apparent that the action of $C_{3}^{+}$ multiplies each addendum of the Bloch sum by $e^{(i4\pi/3)}\equiv e^{(-i2\pi/3)}$, and likewise in general the elements $C_{3}^{\pm}$, $S_{3}^{\pm}$ produce a factor $e^{\mp(i2\pi/3)}$ . 

Now a simple look at the character table of $C_{3h}$~\cite{KDWS} reveals that these complex factors reproduce exactly the characters of the $\Gamma_3$ representation.

The $m=0$ $d-$ like Wannier function, on the other hand, is unaffected by all rotations about the $z$ axis, and it does not affect the overall symmetry of the Bloch function.
%
%
%
%

\begin{figure*}
\includegraphics[width=15cm]{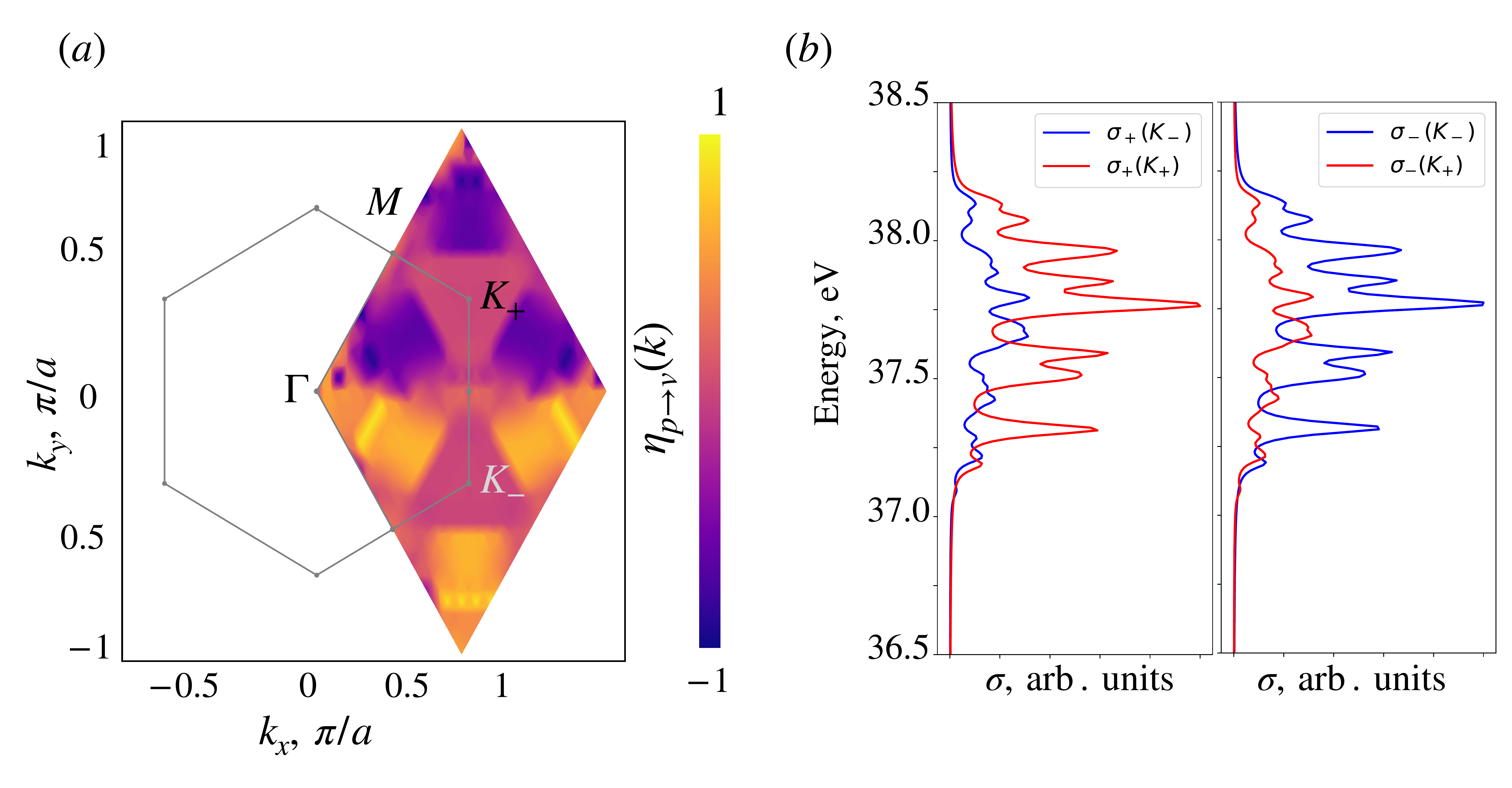}
\caption{Computed normalized degree of optical polarization $\eta$ for transitions from the fourfold degenerate $5_{p_{3/2}}$ levels of tungsten in monolayer $WS_2$ to the lowest conduction band, as a function of quasi-momentum \textbf{k} in the relevant portion of the Brillouin zone (a). Partial absorption cross-section calculated for the circular area around $\bf{K_{\pm}}$ points for left (-) and right (+) circular polarization polarization of the incoming photons (see also Section 3). }
\label{fig:SM-WS2}
\end{figure*}
\section{} 
 
The purpose of this Appendix is to deepen the symmetry assignments of the electronic states at $\mathbf{K_{\pm}}$; for example, it was established, Eq.~(23-24), that the doublet of $p_{1/2}$ states with $j_{z} = \pm 1/2$ belong to the $\Gamma_8$ and $\Gamma_{11}$ representations at $\mathbf{K_+}$; which one corresponds to the $j_{z} = +1/2$ state? This should allow us to identify the wavefunctions for a specific representation and to obtain a direct, intuitive picture of the optical selection rules. 

The procedure to associate a specific wavefunction to an irreducible representation, in the somewhat subtle case of a double group, can be derived from \cite{KDWSf}. 
 
 Let us start from the simple case of the conduction band, where we can basically ignore the spin-orbit interaction. Going back to Eq.~(5) and considering one addendum of the Bloch sum,  we can write the wavefunction for spin-up and -down as
\begin{eqnarray}
e^{i\mathbf{K_+ \cdot R}}\Phi_{2,0}(\mathbf{r} - \mathbf{R}) \left(
\begin{array} {c} 
1  \\   0  
\end{array}
\right),   \ \ \
\\
e^{i\mathbf{K_+ \cdot R}}\Phi_{2,0}(\mathbf{r} - \mathbf{R}) \left(
\begin{array} {c} 
0  \\   1 
\end{array} 
\right).  \ \ \ 
\end{eqnarray}
There are therefore three parts of the wavefunction: Bloch sum, a spherical harmonic type orbital and a spinor. When we act with a $C_{3h}$ operation on it, the wavefunction is multiplied by a factor that is the product of factors related to the Bloch exponential, to the orbital and to the spinor. Evaluating this product gives us the character of the operation in the corresponding irreducible representation, and comparing to the character table we can identify the symmetry type of the wavefunction. The fact that all representations are one-dimensional is of course making this particularly simple. Let us  start with the spin-up case and with $C_{3}^+$. Looking back to Appendix B,  we know already that the Bloch term contributes a factor $e^{-i2\pi/3}$, and the $m=0$ function a factor $1$. The spinors transform according to the $D_{1/2}^{+}$ representation of $O(3)$, and following Eq.~(3.10-3.16) of KDWS, a rotation by an angle $\alpha$ about the $z-$axis corresponds to the matrix
\begin{equation} 
D_{1/2}^{+}(\alpha) = \left[
\begin{array}{cc}
e^{i\alpha/2} & 0 \\
0                    & e^{-i\alpha/2}
\end{array}
\right].
\end{equation}
therefore for a $-2\pi/3$ rotation, there is a factor $e^{-i\pi/3}$ for the spin-up case, $e^{i\pi/3}$ for the spin down. Putting it all together, for $C_{3}^{+}$, we get an overall factor $-1$ for spin up, and $e^{-i\pi/3}$ for spin down. Taking a look at the character table (Table 57 on page 59 of KDWS) this is already sufficient to establish that spin up is $\Gamma_{12}$ and spin down is $\Gamma_{10}$, as Eq.~(10) has already restricted the selection to these two representations. For future reference and to appreciate the care needed for the double groups in the general case, it is instructive to obtain the same result again, but this time choosing the $\sigma_h$ operation. Since Eq.~(8), or the equivalent Eq.~(3-10) and (3-11) of KDWS, lead us to apply the \textit{inverse} of $\sigma_h$, we must avoid the pitfall of identifying $\sigma_h$ as its own inverse: in fact the inverse of $\sigma_h$ is $\overline{\sigma_h}$, which is $\sigma_h$ plus a $2\pi$ rotation (that for a spinor does not coincide with the identity $E$, because it changes all signs). In fact~\cite{KDWS5}:
\begin{equation}
\sigma_{h} \overline{\sigma_h} = E;  \;\;\;\; \sigma_{h}\sigma_{h}= \overline{E}.
\end{equation}
So we know that the effect of $\sigma_h$ on the Bloch factor and on the $m=0$ orbital is a factor of $1$, to find the action on the spinors we must remember that $\sigma_{h} = C_{2} \hat{I}$, a rotation by $\pi$  times the space inversion. We therefore obtain
\begin{equation} 
D_{1/2}^{+} (\overline{\sigma_h}) = -D_{1/2}^{+} (\sigma_{h}) =  \left[
\begin{array}{cc}
-e^{i\pi/2} & 0 \\
0                    & -e^{-i\pi/2}
\end{array}
\right],
\end{equation}
where we used the fact that $D_{1/2}^{+}$ is even under the inversion, and therefore the action of $\sigma_h$ is equivalent to that of $C_2$. We therefore conclude that the effect of $\sigma_h$ for the spin-up state is to multiply it by a factor $-i$, and for the spin-down by $i$, and a look at the character table confirms the $\Gamma_{12}$ assignment for spin up, $\Gamma_{10}$ for spin down.
At $\mathbf{K_-}$ the states transforming as $\Gamma_9$ and $\Gamma_{11}$ correspond, respectively, to spin up and spin down as they are identifiable as the time-reversal partners of, respectively, $\Gamma_{10}$ and $\Gamma_{12}$, as already pointed out in the discussion following Eq.~(11).

Following the reasoning already adopted at the beginning of Section 2B, the same assignments are valid also for $s-$like core levels.

Next we consider the top valence bands. They are unique in their symmetry properties, because as described in Section 2A, the Bloch factor for $\mathbf{K_+}$ and the $m=2$ $d-$orbital combine to form a $C_{3h}$ invariant function, and the same happens at $\mathbf{K_-}$ with the $m=-2$ orbital. Therefore only the behaviour of the up and down spinors characterizes the symmetry behaviour of the top valence bands,  and on the basis of Eq.~(42) we can assign $\Gamma_{8}$ to spin up and $\Gamma_{7}$ to spin down. Again the time reversal invariance imposes $E(\mathbf{K_{+}}, \Gamma_{8})$ =  $E(\mathbf{K_{-}}, \Gamma_{7})$. 
%

Let us now consider the $p_{1/2}$ levels. The corresponding Bloch functions at $\mathbf{K_+}$ can be written, in spinor notation, as:
\begin{equation}
\sum_{j_{1}, j_{2}}e^{i \mathbf{K_+} \cdot \mathbf{R}_{j_{1}, j_{2}}} \Phi_{1/2, j_z} (\mathbf{r} -  \mathbf{R}_{j_{1}, j_{2}}),
\end{equation}
where, for $j_{z} = +1/2$
\begin{equation} 
\Phi_{1/2, 1/2} = \left(
\begin{array}{c}
-\sqrt{1/3} z \\
\sqrt{2/3}(x+iy)
\end{array} 
\right),
\end{equation}
and, for $j_{z} = -1/2$
\begin{equation} 
\Phi_{1/2, -1/2} = \left(
\begin{array}{c}
\sqrt{2/3} (x-iy) \\
-\sqrt{1/3}z
\end{array} 
\right).
\end{equation}
To compute the character of an operation $\hat{C}$, following the usual procedure, we must evaluate
\begin{equation}
\sum_{j_{1}, j_{2}}e^{i \mathbf{K_+} \cdot \hat{C}^{-1} \mathbf{R}_{j_{1}, j_{2}}} D_{1/2}^{-}(\hat{C}^{-1})\Phi_{1/2, j_z} (\hat{C}^{-1}(\mathbf{r} -  \mathbf{R}_{j_{1}, j_{2}}).
\end{equation}

The effect on the exponential Bloch factor can be evaluated as in the previous sections, and for the spinor we apply the procedure described by KDWS on p. 10 ff., Eq (3-13) to (3-16). For example, for the $C_{3}^+$ operation, the exponential acquires a factor $e^{-i2\pi/3}$, for the $D$ matrix we can use Eq.~(C3) (the $\pm$ parity sign intervening for improper rotations only) 
\begin{equation}
D_{1/2}^{-}(C_{3}^{-1}) = \left[
\begin{array}{cc}
e^{-i\pi/3} & 0 \\
0                    & e^{i\pi/3}
\end{array}
\right]
\end{equation}
and for the spinor component we simply apply $C_{3}^{-}(z) = z$,  $C_{3}^{-}(x\pm iy) = e^{i\mp2\pi/3}(x\pm iy)$, to obtain
\begin{widetext}
\begin{equation}
C_{3}^{+} \sum_{j_{1}, j_{2}}e^{i \mathbf{K_+} \cdot \mathbf{R}_{j_{1}, j_{2}}} \left(
\begin{array}{c}
-\sqrt{1/3} z \\
\sqrt{2/3}(x+iy)
\end{array} 
\right)=
e^{-i2\pi/3}\sum_{j_{1}, j_{2}}e^{i \mathbf{K_+} \cdot \mathbf{R}_{j_{1}, j_{2}}} \\  \left(
\begin{array}{c}
-e^{-i\pi/3} \sqrt{1/3} z \\
e^{i\pi/3} \sqrt{2/3}e^{-i2\pi/3}(x+iy)
\end{array} 
\right).
\end{equation}
\end{widetext}
Therefore, the $1/2,1/2$ partner of the $p_{1/2}$ doublet is multiplied by $-1$ when acted upon by $C_3^{+}$, which associates it to the $\Gamma_{11}$ representation (see Eq.~(23)), while, necessarily, the $1/2,-1/2$ partner belongs to $\Gamma_{8}$.  

The assignment at $\mathbf{K_-}$ can be inferred by time reversal symmetry, with $1/2,1/2$ belonging to $\Gamma_7$, and $1/2,-1/2$ belonging to $\Gamma_{12}$.

A similar treatment yields the assignments for the four  irreducible representations for the $p_{3/2}$ quartet. The four Wannier functions can be expressed in spinor notation as
\begin{equation} 
\Phi_{3/2, j_{z}=3/2} =   \left(
\begin{array}{c}
(x+iy) \\
0
\end{array} 
\right),
\end{equation}

\begin{equation} 
\Phi_{3/2, j_{z}=1/2} =   \left(
\begin{array}{c}
\sqrt{2/3}z \\
\sqrt{1/3} (x+iy)
\end{array} 
\right),
\end{equation}

\begin{equation} 
\Phi_{3/2, j_{z}=-1/2} =   \left(
\begin{array}{c}
\sqrt{1/3}(x-iy)\\
\sqrt{2/3}z
\end{array} 
\right),
\end{equation}

\begin{equation} 
\Phi_{3/2, j_{z}=-3/2} =   \left(
\begin{array}{c}
0 \\
 (x-iy)
\end{array} 
\right)
\end{equation}
and repeating the previous procedure we get \\
at $\mathbf{K_+}$:
\begin{widetext}
\begin{equation} 
\Phi_{3/2, j_{z}=3/2} \longrightarrow \Gamma_{9} \;\;\;\; \Phi_{3/2, j_{z}=1/2} \longrightarrow \Gamma_{11} \;\;\;\; \Phi_{3/2, j_{z}=-1/2} \longrightarrow \Gamma_{8} \;\;\;\; \Phi_{3/2, j_{z}=-3/2} \longrightarrow \Gamma_{7} 
\end{equation}
and, by time-reversal at  $\mathbf{K_-}$:
\begin{equation} 
\Phi_{3/2, j_{z}=3/2} \longrightarrow \Gamma_{8} \;\;\;\; \Phi_{3/2, j_{z}=1/2} \longrightarrow \Gamma_{7} \;\;\;\; \Phi_{3/2, j_{z}=-1/2} \longrightarrow \Gamma_{12} \;\;\;\; \Phi_{3/2, j_{z}=-3/2} \longrightarrow \Gamma_{10}.
\end{equation}
\end{widetext}

\section{\texorpdfstring{$WS_2$}{Lg}: core-to-conduction band transitions} 
 
For the experiment described in Section 4, it is essential to have valley selectivity for the transitions from and to the top of the valence band. However, the valley selectivity is not expected for transitions from $p-$ core levels to the conduction band minima at the $\bf{K}_{\pm}$ points, according to the findings of the group theoretical analysis of Section 2. It is interesting to explore the first principle calculations for the $WS_2$ monolayer in some detail (see Figure~\ref{fig:SM-WS2}). In fact, transitions from each individual $W$ $5p_{3/2,m_{j}}$ level are valley selective; but transitions with positive and negative $m_j$ have an equal and opposite valley sensitivity and cancel each other completely around the $\bf{K}_{\pm}$ points. Similarly to what is found also in $MoS_2$, there are small secondary valleys in between $\bf{\Gamma}$ and $\bf{K_{\pm}}$ points which show a partial degree of dichroism  (see Figure~\ref{fig:SM-WS2}(b)). The latter could be relevant for the future studies involving relaxation and intervalley scattering. 
\section{X-ray edges for \texorpdfstring{$Mo$}{Lg} and \texorpdfstring{$W$}{Lg}}
\label{appendix:table}

\onecolumngrid
\begin{widetext}
\begin{table}[h]
\caption{\textit{Atomic binding energies of $Mo$ and $W$  core levels in $eV$; hard x-ray levels $>3\, keV$ are omitted} (from ref.~\cite{Berkeley})}
\begin{tabular*}{\textwidth}{c |@{\extracolsep{\fill}} ccccccc}
\hline
\hline
Level & $L_{1}(2s_{1/2})$ & $M_{1}(3s_{1/2})$ & $N_{1}(4s_{1/2})$ & $O_{1}(5s_{1/2}) $ & $M_{2}(3p_{1/2})$ & $M_{3}(3p_{3/2})$ & $M_{4}(3d_{3/2})$ \\
\hline
\hline
Mo & 2866 & 506.3 & 63.2 &   &  411.6& 394.0 & 231.1   \\
\hline
W &  & 2820 & 594.1 &  75.6 & 2575 & 2281 & 1872  \\
\hline
\hline
Level &  $M_{5}(3d_{5/2})$& $N_{2}(4p_{1/2})$  &  $N_{3}(4p_{3/2})$ & $N_{4}(4d_{3/2})$ & $N_{5}(4d_{5/2})$  & $O_{2} (5p_{1/2})$ & $O_{3} (5p_{3/2})$\\
\hline 
\hline
Mo & 227.9 & 37.6 & 35.5 &  &  \\
\hline
W &  1809& 490.4 &423.6 & 255.9 & 243.5 & 45.3 & 36.8 \\
\end{tabular*}
\end{table}
\end{widetext}
\twocolumngrid

\footnotetext{Table 62 on page 62 of KDWS}


\begin{thebibliography}{99}
\bibitem{KDWS-t5759} The $C_{3h}$ character table is Table 57, page 59 of ref. \cite{KDWS}.
\end{thebibliography}
\end{document}